\documentclass[fleqn,aip,jcp,preprint,a4paper,onecolumn]{revtex4}

\usepackage[]{amsmath}
\usepackage{amssymb}
\usepackage[dvips]{graphicx}
\usepackage{color}
\usepackage{tabularx}
\usepackage{mathtools}
\usepackage{algpseudocode}
\usepackage{enumitem}
\usepackage{multirow}
\usepackage{epstopdf}  
\usepackage{bm}
\usepackage{footmisc}
\usepackage{cleveref}


\newcommand{\recheck}[1]{{\color{black} #1}}

\newcommand{\dpsr}{\mathrm{sr}}

\newcommand{\rec}{\mathrm{rec}}

\newcommand{\dft}{\mathrm{DFT}}
\newcommand{\dplrc}{\mathrm{DPLR}}
\newcommand{\fitting}{\mathcal{F}}
\newcommand{\descrpt}{\mathcal{D}}
\newcommand{\environ}{\mathcal{R}}
\newcommand{\embedng}{\mathcal{G}}
\newcommand*\chem[1]{\ensuremath{\mathrm{#1}}}
\newcommand{\cwcdisp}{D}

\newcommand*{\citen}[1]{%
  \begingroup
    \romannumeral-`\x % remove space at the beginning of \setcitestyle
    \setcitestyle{numbers}%
    \cite{#1}%
  \endgroup   
}

\begin{document}

\title{
 A deep potential model with long-range electrostatic interactions
  }
\author{Linfeng Zhang}
% \affiliation{Program in Applied and Computational Mathematics, Princeton University, Princeton, NJ 08544, USA}
\affiliation{DP Technology, Beijing, People’s Republic of China.}
% \author{Jiequn Han}
% \affiliation{Program in Applied and Computational Mathematics, 
% Princeton University, Princeton, NJ 08544, USA}
\author{Han Wang}
\email{wang\_han@iapcm.ac.cn}
\affiliation{Laboratory of Computational Physics,
  Institute of Applied Physics and Computational Mathematics, Fenghao East Road 2, Beijing 100094, P.R.~China}

\affiliation{HEDPS, CAPT, College of Engineering, Peking University, Beijing 100871, P.R.~China}  
  
\author{Maria Carolina Muniz}
\affiliation{Department of Chemical and Biological Engineering, Princeton University, Princeton, New Jersey 08544, USA
}
\author{Athanassios Z Panagiotopoulos}
\affiliation{
Department of Chemical and Biological Engineering, Princeton University, Princeton, New Jersey 08544, USA
}
\author{Roberto Car}
\email{car@princeton.edu}
\affiliation{Department of Chemistry, Department of Physics, Program in Applied and Computational Mathematics, Princeton Institute for the Science
and Technology of Materials, Princeton University, Princeton, New Jersey 08544, USA}

\author{Weinan E}
% \email{weinan@math.princeton.edu}
\affiliation{School of Mathematical Sciences, Peking University, Beijing 100871, P.R.~China}
\affiliation{AI for Science Institute, Beijing, P.R.~China}
\affiliation{Department of Mathematics and Program in Applied and Computational Mathematics, Princeton University, Princeton, NJ 08544, USA }

\begin{abstract}
Machine learning models for the potential energy of multi-atomic systems, such as the deep potential (DP) model, make possible molecular simulations with the accuracy of quantum mechanical density functional theory, at a cost only moderately higher than that of empirical force fields. However, the majority of these models lack explicit long-range interactions and fail to describe properties that derive from the Coulombic tail of the forces. To overcome this limitation we extend the DP model by approximating the long-range electrostatic interaction between ions (nuclei+core electrons) and valence electrons with that of distributions of spherical Gaussian charges located at ionic and electronic sites. The latter are rigorously defined in terms of the centers of the maximally localized Wannier distributions, whose dependence on the local atomic environment is modeled accurately by a deep neural network. In the deep potential long-range (DPLR) model, the electrostatic energy of the Gaussian charge system is added to short-range interactions that are represented as in the standard DP model. The resulting potential energy surface is smooth and possesses analytical forces and virial. Missing effects in the standard DP scheme are recovered, improving on accuracy and predictive power. By including long-range electrostatics, DPLR correctly extrapolates to large systems the potential energy surface learned from quantum mechanical calculations on smaller systems. We illustrate the approach with three examples, the potential energy profile of the water dimer, the free energy of interaction of a water molecule with a liquid water slab, and the phonon dispersion curves of the NaCl crystal.

\end{abstract}

\maketitle

\setlength{\parskip}{.5em}

\section{Introduction}

% \WH{success of the ML PES.}

Modeling materials in the electronic ground-state requires an accurate description of the Born-Oppenheimer potential energy surface (PES). First-principles models based on density functional theory (DFT) have good predictive power, but high computational complexity. Empirical force fields are more efficient but often less accurate than DFT, and are typically not applicable when chemical bonds change within a system of interest. 
PES models based on machine-learning (ML) offer a promising route to overcome the dilemma of accuracy versus efficiency~\cite{behler2007generalized,bartok2010gaussian,rupp2012fast,bartok2013representing,thompson2015spectral,shapeev2016moment,schutt2017quantum,smith2017ani,chmiela2017machine,chmiela2018towards,yao2018tensormol,han2018deep,zhang2018deep,zhang2018end,lubbers2018hierarchical,unke2019physnet}. 
These models combine force field efficiency with first-principles accuracy and have been successful 
in a number of applications that could not be tackled with direct
first principles studies. 
Here, we focus on one such method, the deep potential (DP) scheme~\cite{zhang2018deep,zhang2018end}, which, upon training with DFT data, can achieve 
accuracy comparable to that of \textit{ab initio} molecular dynamics (AIMD). On high-performance computer platforms, deep potential molecular dynamics (DPMD) simulations can model millions of atoms on timescales of tens of nanoseconds or even longer~\cite{lu202186,jia2020pushing}. 
Successful applications to date include studies of
the signatures of a liquid-liquid phase transition in supercooled water~\cite{gartner2020signatures}, nucleation of crystalline Si from the melt~\cite{bonati2018silicon}, 
phase diagrams of Ga~\cite{niu2020ab} and  water~\cite{zhang2021phase}, 
structural motifs in the growth of quasicrystals~\cite{han2020dynamic}, the reactive uptake of \chem{N_2O_5} in interfacial processes~\cite{galib2021reactive},
reactive processes in combustion~\cite{zeng2020complex}, 
the dissociative chemisorption of water at the water-titania interface~\cite{andrade2020free}, and of the 1D cooperative diffusion in a simple cubic crystal~\cite{wang2021electronically}.

%\RC Here we should give a few examples
%DP helps boosting the efficiency while maintaining accuracy in many applications, for example, it is used to provide evidence on the longly-debated %problem of liquid-liquid phase transition in supercooled water~\cite{}, is used to investigate the nucleation in Silicon at the accuracy of DFT with %SCAN~\cite{} exchange-correlation function~\cite{}, is to chart the phase digram of Gallium~\cite{}, and is used to detect the structure motif in %the growth of quasicrystal in solid-state transformations~\cite{}.

% \WH{Assumption of near-sightness, counter examples.}
Most ML models, including DP, assume that the PES can be represented by a sum of atomic terms describing the many-body interactions among the atoms within some fixed cutoff distance. This construction is computationally advantageous as its complexity scales linearly with system size. On physical grounds, however, it entails a severe approximation, because electrons and nuclei are charged particles that experience long-range Coulomb forces. In spite of that, local effective interaction models work well in many cases. In metals, this can be attributed to complete electronic screening. In insulators, screening is incomplete, yet, the interatomic interactions are often represented
satisfactorily by short-range ML models. These models are trained on DFT data, which include long-range electrostatics, suggesting
that some features of the long-range interactions may be mimicked by short-range models. 
While this treatment is sufficient in many cases, particularly when dealing with systems that are, on average, homogeneous and neutral at the molecular scale, failures of the short-range models in cluster, interfacial and vapor phase properties are well known~\cite{yue2021short,niblett2021learning}.
A quintessential example of a property
that cannot be simulated by short-range models is the longitudinal-transverse splitting of the long-wavelength
optical phonon modes in polar crystals. Long-range electrostatics should be even more important in heterogeneous systems, where 
charge separation at large scale may be induced by external fields or by chemical potentials. Examples include electrolyte solutions 
in contact with electrodes~\cite{scalfi2021molecular},
%The validity of the assumption relies on the near-sightness of the electronic matter~\cite{}. 
%In conducting system, this would be a reasonable approximation, because the perturbation to the electronic structure is near-sightness due to the %screening effect of the metallic electrons.
%However, the assumption may fail in insulating systems, where electrostatic potential is long-ranged and plays a critical role. 
%For example, 
dipolar surfaces~\cite{patra2003molecular,wohlert2004range}, protein folding problems~\cite{zhou2018electrostatic}, and the binding affinity of ligands to proteins~\cite{rocklin2013calculating}. 

% However, the assumption may fail in insulating systems, where electrostatic potential is long-ranged. 
% This usually does not introduce significant error in MLP simulations~\cite{morawietz2016van}. 
% The possible reason is that the system is homogeneous and the change and dipole are locally and statistically neutral, thus the long-range interactions are canceled in the sense of statistical average~\cite{}.
% However, long-range interaction plays a critical role in modeling many systems, in which a local neutrality cannot be assumed, for example, dipolar surfaces~\cite{patra2003molecular,wohlert2004range}, protein folding problems~\cite{zhou2018electrostatic}, and the binding affinity of ligands to proteins~\cite{rocklin2013calculating}.

% \WH{incouperating long-range interaction in literature}
A possible way of incorporating long-range electrostatics within ML models is to assume 
that the PES has two contributions. One accounts 
for short-range interactions and is constructed as in standard ML models. The other accounts for
long-range interactions and is approximated by the electrostatic energy of a system of partial point charges located at the atom sites, 
with the added constraint of global charge conservation. 
The effective charges can be either defined empirically, as commonly done in the context of 
force-field models ~\cite{deng2019electrostatic}, or can be found by machine learning~\cite{artrith2011high,bereau2015transferable,yao2018tensormol,bereau2018non,unke2019physnet},
in which case the partial charges are matched to reference charges extracted from DFT calculations by
partitioning the electronic charge density among the atoms~\cite{artrith2011high,bereau2015transferable,bereau2018non}. 
A difficulty with this approach is that an atom in a molecule, or a material, does not have a well defined charge, because the charge densities of neighboring atoms overlap.
As a consequence, different partitioning schemes may lead to different results~\cite{nebgen2018transferable,sifain2018discovering}. 
To reduce this ambiguity, charge partitioning schemes have been combined with the concepts of electronegativity and hardness~\cite{ghasemi2015interatomic}, 
which, however, also lack a rigorous definition. 
Interestingly, the accuracy and transferability of molecular ML models with effective atomic charges 
were found to improve significantly when the molecular dipole moment was added to energy and forces in the training data~\cite{ghasemi2015interatomic,ko2021fourth,ko2021general}. 
\recheck{This condition is implemented in PhysNet~\cite{unke2019physnet}, a model in which the partial charges are trained to best reproduce total energy and dipole of a system.}
% It is noted that the concept of total dipole  is ill-defined for ionized condensed systems under periodic boundary conditions.
% In this context, it is important to describe the dipolar fluctuations,  which are not possible to investigate with the partial atomic charges~\cite{vanderbilt2018berry}. 
A more flexible representation than the point charge models 
is provided by the long-distance equivariant (LODE) framework, 
in which the charge density is approximated by a sum
of atom centered spherical Gaussians, or other localized functions~\cite{grisafi2019incorporating,grisafi2021multi}.
LODE successfully describes the mutual interaction of charged molecular species at large separation, a 
property beyond the reach of short-range ML models.
   
The above schemes can be seen as different realizations of a coarse-graining transformation that approximates   
the electronic density with a sum of atom centered spherical contributions. 
% This representation can provide a good overall fit of the charge density, but 
% %does
% %not capture the quantum mechanical character of the covalent bonds, %and, in condensed phase,
% does not describe rigorously the dipole moment and the fluctuations of the polarization that couples with the electric field generated by distant charges.
\recheck{This representation can provide a good overall fit of the charge density and of the total dipole of a molecule, but, in general, does not describe correctly the dipolar fluctuations that couple with the electric field generated by distant charges in extended systems within periodic boundary conditions.
The reason is that the dipole of a periodic crystal can only be defined modulo a quantum, and the polarization fluctuations typically include dipolar fluxes across the cell boundary, originating from the delocalized nature of the quantum mechanical electronic charge distribution. 
These fluxes are related to the phase of the wavefunction and cannot be described, in general, by partial atomic charges (see e.g., Ref.~\citen{vanderbilt2018berry}). 
Interestingly, an exception to this rule occurs when a crystal is made of nonoverlapping molecular units, because, in this case, the total dipole is the sum of the molecular dipoles.}

In this work, we propose an alternative deep learning model for insulating systems that overcomes this limitation.
As in previous approaches, we assume that the PES has short- and long-range contributions. 
The short-range contribution has the standard form of the DP model. The long-range contribution is the electrostatic energy of a system of spherical Gaussian charges, associated with the ions (nuclei+frozen core electrons) and the valence electrons, located, respectively, at the ionic and electronic sites. The latter are defined by the averages of the positions of the maximally localized Wannier centers~\cite{marzari2012maximally} associated with specific atoms~\cite{zhang2020deep}. 
In general, ionic and electronic sites do not coincide. The electronic site coordinates are not independent dynamical variables, but are fixed by the atomic configuration, as required by the Born-Oppenheimer adiabatic separation of ionic and electronic dynamics.
% \recheck{As an approximation, we assume that the electronic sites, hereafter also called Wannier centroids (WCs), depend on the atoms in their neighborhood, thus the polarization due to other part of the system outside the neighborhood is ignored.}
% Due to the nearsightedness of the electronic matter~\cite{prodan2005nearsightedness}, the electronic sites, hereafter also called Wannier centroids (WCs), depend only on the atoms in their neighborhood.
\recheck{We assume that the electronic sites, hereafter also called Wannier centroids, depend only on the atoms in their neighborhood, a conjecture that can be rationalized in terms of the so-called nearsightedness of electronic matter~\cite{prodan2005nearsightedness}. 
We found that this condition is well satisfied in all systems we have studied to date.}
Thus, the environmental dependence of the centroids can be modeled accurately by a deep neural network like deep Wannier (DW), 
which was introduced in Ref.~\citen{zhang2020deep} to describe the dielectric response of insulators.  
%\LZ{two ``model'' appears in this sentence, which seems to be a bit confusing}
By combining DP and DW we construct a model of the PES that includes explicit long-range electrostatic interactions. 
The new scheme, called deep potential long-range (DPLR), has several advantages relative to atom centered models. 
First, the centroid distributions have integer charges and automatically conserve the total charge of the system. Second,
their positions derive from a unitary transformation in the subspace of the occupied Kohn-Sham orbitals. 
%Thus, the electron 
%density in insulators is rigorously \recheck{approximated} by a sum of %localized distributions centered at the Wannier centroids. 
% By construction, this representation 
% %captures the covalent character of the chemical bonds, and
% describes rigorously molecular dipoles in finite systems and
% polarization fluctuations in condensed media.
\recheck{By construction, this representation rigorously describes molecular dipoles in finite systems and dipolar fluctuations in condensed media (see e.g.~Ref.~\citen{vanderbilt2018berry}).}
The spherical Gaussian distributions adopted in DPLR differ from the maximally localized Wannier distributions in the quadrupole and higher moments. This results in a potential energy error that converges well with size. In contrast,
the errors of atom centered distributions start at the dipole level, and the potential energy error is only conditionally convergent with 
size. 
%In principle, it would be possible to have errors starting at the %octupole level by using
%non-spherical Gaussians that match the spreads of the Wannier %distributions.   
%The environmental dependence of the spreads could be learned by a %suitable generalization of DW, but the ensuing increase of complexity %and computational cost seems unnecessary on the basis of tests with %spherical Gaussians.
The DPLR model is physically 
meaningful, symmetry preserving, and smooth, so that forces and virial can be analytically calculated, preserving the conservative properties of the adopted molecular dynamics scheme. 
The approach is also computationally efficient, as fast algorithms 
for calculating Ewald sums for Gaussian charges under periodic boundary conditions are well established~\cite{ewald1921die,hockney1988computer,darden1993pme,essmann1995spm}. DPLR describes polarization fluctuations in finite and extended systems. Thus, it also allows us to model the response of a system to an externally applied field. 

%within the linear regime.    

While preparing this manuscript, we became aware of a
recent preprint by Gao and Remsing~\cite{gao2021self}, in which the
electronic structure information encoded in the Wannier centers is used to construct a ML model of the PES including long-range electrostatics. This model, called self-consistent field neural network (SCFNN), differs from DPLR because the separation between short- and long-range
contributions is not done in the way in which the training data are handled, but in the way in which they are generated. 
The model requires short-range DFT data obtained, in principle, from calculations with a truncated Coulomb potential. 
A drawback of this formulation is that it introduces a self-consistency condition for the Wannier center positions. 
As a consequence, the simplicity of the ML construction is lost, and the dynamics of the model is no longer conservative, unless
formulations such as those proposed by Car and Parrinello are introduced~\cite{car1985unified}. In practice, separating standard DFT data into data
for a truncated Coulomb potential and data for its long-range counterpart is not straightforward, and the authors overcome this difficulty by invoking linear response conditions. 
%This suggests that the complication of self-consistency may be unnecessary, as is the case with DPLR.
%\AZP{I find this sentence somewhat unclear.RC: I agree and dropped the sentence, which seems unnecessary to me, as we return on the issue in the Conclusions, while here we have already stated clearly that DPLR does not require self-consistency}

In this manuscript, we focus on DPLR and
how it compares to the original DP model.
Theory and method are presented in
Sect.~II. In Sect.~III we construct the DP, DW, and DPLR models for water that we use in the applications discussed in Sect.~IV.
In that section we compare DPLR, DP, and direct DFT calculations for the binding energy curve of the water dimer and for the free energy profile of a water molecule as a function of the distance from a liquid water slab.
In Sect.~V we construct DP and DPLR models for small atomic displacements around the classical ground-state in crystalline NaCl, and use these models to compute the phonon dispersion curves.
We show that the non-analytic contribution to the dynamical matrix, arising from long-range dipole-dipole interactions, can be calculated with DW and its polarizability extension~\cite{sommers2020raman}. 
We then study the evolution of the analytical phonon dispersion curves calculated with finite supercells of increasing size. 
While the modes calculated with the DP model do not exhibit size dependence, those calculated with DPLR depend on size, and partially recover the longitudinal-transverse splitting of the optical modes, within the limitations of the finite supercells used in the calculations. \recheck{In Sect. VI we discuss how the model can be used to study the response of a system to an externally applied electric field.}  Finally, Sect.~VII is devoted to our conclusions.  

% To model the long-range contribution, the system electron density is approximated by a Gaussian distribution centered at the Wannier centoids. 
%Then the long-range contribution is approximated by the
%electrostatic interaction between Gaussian charge distributions centered at the position of ions and Wannier centroids. 
%We prove that the error introduced by such an approximation is controlled by a dipole-quadrupole interaction term, and rapidly converges.
%This energy model is physically meaningful, symmetry preserving and smooth so that the force and pressure are analytically calculated.
%The approach is also computationally efficient, as the fast alogrithms for calculate a group of Gaussian charges under periodic boundary condition %is well established~\cite{ewald1921die,hockney1988computer,darden1993pme,essmann1995spm}.

\section{Theory and method}

We focus on extended systems modelled with a periodically repeated supercell. Finite systems can be investigated in this way, provided the supercell is large enough that interactions between periodic images can be neglected. With periodic boundary conditions the system must be electrically neutral. However, the scheme could be easily extended to different choices of boundary conditions.     

\subsection{Electrostatic energy}

\begin{figure}
    \centering
    \includegraphics[width=0.4\textwidth]{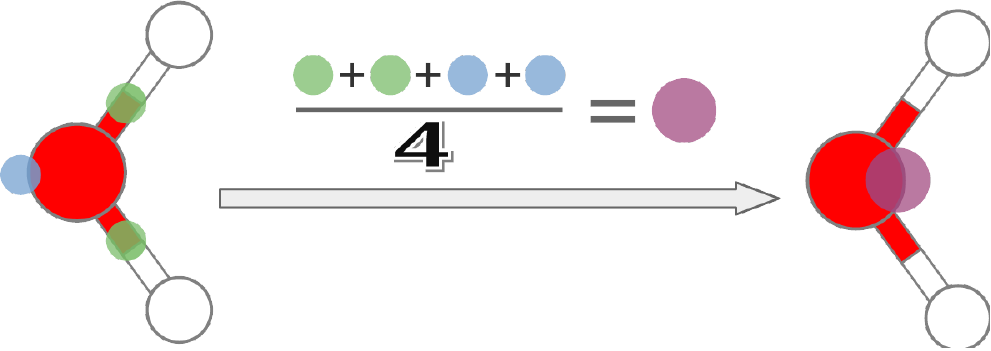}
    \caption{The Wanner centroid (WC) associated with a water molecule is represented by the purple dot. The maximally localized Wannier Centers (MLWCs) corresponding to bond pairs and lone pairs are represented by green and blue dots, respectively. The WC is the geometric center of the 4 MLWCs. In molecular dynamics trajectories of water systems the WCs are uniquely associated with the oxygen atoms.}
    \label{fig:wc}
\end{figure}

In the reference DFT model we adopt a pseudopotential framework. Thus, in what follows, electrons stand for valence electrons, and ions for nuclei plus frozen core electrons.
However, the method would be applicable to all-electron DFT methods as well, in which case ions would stand for nuclei and electrons would include valence and core. 
%\recheck{The generalization to all-electron DFT methods would be possible.}
% All electron generalizations would be possible. 
We indicate the charge density of ions and electrons by $\rho_{\textrm{ion}}(r)$ and $\rho_e(r)$, respectively. When the system is in the ionic configuration
$\{R_I\}$, $\rho_{\textrm{ion}}(r)$ is given by: 
\begin{align}
    \rho_{\textrm{ion}}(r) = \sum_I \rho_I (r-R_I),
\end{align}
where $R_I$ indicates ion coordinates, and $\rho_I (r) = q_I \delta (r)$, in terms of $q_I$, the charge of the ion $I$, and of $\delta (r)$, the Dirac delta distribution. In insulators, the density of the electrons, $\rho_e(r)$, can be represented by a sum of local distributions 
via a unitary transformation of the occupied Bloch orbitals onto maximally localized Wannier functions~\cite{marzari2012maximally}. The centers of the local distributions are called maximally localized Wannier centers (MLWC). 
\recheck{Often, during molecular evolution}, the same MLWCs can be uniquely
assigned to the same atom along a trajectory.
For example, in water systems the same four MLWCs are always nearest 
to the same oxygen atom irrespective of liquid diffusion and molecular dissociation. It is then convenient to define the $n$-th Wannier centroid (WC), $W_n$, as the average of
the MLWCs assigned to a given atom, as illustrated in Fig.~\ref{fig:wc} for a water molecule. 
\recheck{Lumping into a single centroid associated to the O atom the four Wannier centers of a water molecule has the following advantages: (i) it makes straightforward to satisfy, within the deep neural network (DNN) model, the Born-Oppenheimer constraint of parametric dependence of the ground-state electronic properties on the atomic coordinates; (ii) it eliminates the need for imposing permutational symmetry between equivalent Wannier centers in a molecule; (iii) it is computationally efficient.
We note, however, that a unique association of the Wannier centers to specific atoms is not possible in presence of electron transfer between different atoms. Dealing with these events requires a significant generalization of the method. Ignoring this possibility,}
$\rho_e (r)$ takes the form:           
%$\rho_e(r)$. 
%Within DFT this distribution is known but in our ML model we only know the ion coordinates 
%we only know the ion coordinate within DP and the maximally localized Wannier centers (MLWC) and their centroid, which is defined by grouping the %MLWCs associating to the same atom~\cite{zhang2020deep}, within DW.
%Now we denote the centroid of Wannier centers (CWC) by $W_n$. The unknown distribution $\rho_e(r)$ can be represented by a sum over localized %distributions $\rho_n$ via a unitary tranform, i.e.
\begin{align}
    \rho_e(r) = \sum_n \rho_n(r-W_n),
\end{align}
where $\rho_n$ is the sum of the maximally localized Wannier distributions associated with the $n$-th WC. 
For simplicity, we consider
spin saturated systems, in which each maximally localized Wannier distribution carries a charge of 2 electrons ($-2e$). Thus, $\rho_n$ carries an integer charge $q_n$, equal to $-2e$ times the number of Wannier functions associated with the $n$-th centroid. For instance, in the water examples of the manuscript, $q_n = -8e$. The charge neutrality condition implies that the total charge density, i.e., $\rho_t (r) = \rho_{\textrm{ion}} (r) + \rho_e (r)$, satisfies:
%The distribution $\rho_n$ is unknown to DPMD, but
%we know two things about it: (1) it is localized and its center is located at $W_n$, (2) if its integral is equal to $q_n$ then $\sum_I q_I + \sum_n %q_n = 0$, i.e., global charge neutrality is guaranteed. Thus, we have for the total charge distribution $\rho_t(r)$
\begin{align}
    \int dr \rho_t(r) = \int dr ( \rho_{\textrm{ion}} (r) + \rho_e (r) ) = 0
\end{align}
Thus, the total electrostatic energy is well defined and given by:
\begin{align}\label{eqn:total-e}
    E_t = \frac 12 \int dr dr' \rho_t(r) v(r-r') \rho_t(r')
\end{align}
Here $v(r) = 1/\vert r\vert$ is the Coulomb interaction, and it is understood that the (infinite) self-energy of the ionic point charges should be subtracted from the integral. The energy in Eq.~\eqref{eqn:total-e}  is calculated exactly in the DFT reference model, while it is approximated, in the DPLR model, by the electrostatic energy of a system of spherical Gaussian charge distributions $g_I$ and $g_n$ located at ionic ($R_I$) and WC ($W_n$) sites, respectively. In DPLR one adds to this energy a short-range contribution described by the DP part of the model, which takes care of all the additional many-body effects within $r_c$, the cutoff radius of the model. Thus, all residual errors of the approximation for the electrostatic energy should be associated with long-range effects, for which $r > r_c$.    
The Gaussian distributions $g_I$ and $g_n$ integrate to $q_I$ and $q_n$, respectively, but are assumed to have the same spread $1 / (\sqrt 2 \beta)$, with $\beta$ an adjustable parameter. 
% guassian: \exp ( - \beta^2 r^2 )
The true distributions, $\rho_I$ and $\rho_n$, differ from $g_I$ and $g_n$ according to: 
\begin{align}\label{eqn:gi-di}
    &\rho_I(r) = g_I(r) + \rho_I(r) - g_I(r) = g_I(r) + \Delta_I(r)\\\label{eqn:gn-dn}
    &\rho_n(r) = g_n(r) + \rho_n(r) - g_n(r) = g_n(r) + \Delta_n(r)
\end{align}
The distributions $\Delta_I$ and $\Delta_n$, in Eqs.~\eqref{eqn:gi-di} and ~\eqref{eqn:gn-dn}, integrate to zero. $\Delta_I$ is known, having all its mutipole moments equal to zero, while $\Delta_n$ has zero dipole moment, but is otherwise unknown. We choose the spread parameter $\beta$ so that the Gaussian distributions are at the same time smooth on the atomic scale and 
well localized within $r_c$. More details on the choice of $\beta$ will be given when discussing specific examples later in the manuscript.
While the electrostatic potential generated by $\Delta_I$, which has spherical symmetry, decays to 0 exponentially for $r > r_c$, the potential generated by $\Delta_n$ decays to 0 algebraically for $r > r_c$ because, in general, the distribution $\Delta_n$ has non-zero quadrupole and higher multipole moments. Thus, the potential generated by $\Delta_n$ decays at least as fast as $r^{-3}$ for $r > r_c$ and has magnitude controlled by the quadrupole moments specific to the system. Typically, we expect quadrupoles on the order of 1 a.u., which would correspond to $\Delta_n(r=r_c)$ being approximately equal to 1 meV for $r_c = 6$ A. 
%Finally, the long-range electrostatic contribution to Eq. \eqref{eqn:total-e} is calculated with $g_I$ and $g_n$ in place of $\rho_I$ and $\rho_n$.
Given that the DP part of the model takes care of the effects within $r_c$, the error of the above procedure is associated with effects 
beyond $r_c$, which are controlled primarily by $\Delta_n$. 

The total ionic and electronic Gaussian distributions are given by:
\begin{align}
    & G_{\textrm{ion}}(r) = \sum_I g_I(r-R_I)\\
    & G_e(r) = \sum_n g_n (r-W_n)
\end{align}
The  corresponding total Gaussian charge distribution is $G_t(r) = G_{\textrm{ion}}(r) + G_e(r)$. 
The electrostatic energy associated with $G_t$ is easily calculated in Fourier space~\cite{ewald1921die}
\begin{align}\label{eqn:e-dplr}
    E_{G_t} & = \frac{1}{2\pi V} \sum_{m\neq 0, \vert m\vert \leq L} \frac{\exp(-\pi^2 m^2/\beta^2)}{m^2}  S^2(m),
\end{align}
%\RC{In the above Equation M is used for the maximum value of m, however M has also been used earlier to indicate cell dipole}
%\MCM{how is the beta parameter obtained?} \WH{It is the inverse of the Gaussian spread, explained below \eqref{eqn:sm}.}
where $L$ is the cutoff in Fourier space, and $S(m)$, the structure factor, is given by:
\begin{align}\label{eqn:sm}
    S(m) = \sum_I q_I e^{-2\pi i m  R_I} + \sum_n q_n e^{-2\pi i m W_n}
\end{align}
%In Eq.~\eqref{eqn:e-dplr} we assumed that the Gaussian distributions $g_I$ and $g_n$ have the same spread $\sqrt 2/\beta$.
Due to the smoothness of the distributions, the Fourier sum in Eq.~\eqref{eqn:e-dplr} converges rapidly. 
Fast algorithms for calculating the electrostatic energy are available, such as the smooth particle mesh Ewald (SPME)~\cite{darden1993pme,essmann1995spm} and the particle-particle-particle-mesh (PPPM)~\cite{hockney1988computer} methods. The latter is adopted in the applications discussed in this manuscript.
The electrostatic energy $E_{G_t}$ in Eq. \eqref{eqn:e-dplr} contains also short-range contributions. To avoid double counting, $E_{G_t}$ is subtracted from the total potential energy when training the DP part of the DPLR model.

The error $\mathcal E$, due to neglecting the contribution of $\Delta_n$ to the electrostatic energy, is the sum of two terms, $\mathcal E_1$ and $\mathcal E_2$, given respectively by:
\begin{align}\label{eqn:error-e1}
    &\mathcal E_1 = \int drdr' G_t(r) v(r-r') \sum_n\Delta_n (r') \\ \label{eqn:error-e2}
    &\mathcal E_2 = \int drdr' \sum_m\Delta_m(r) v(r-r') \sum_n\Delta_n(r)
\end{align}
The two integrals above are well defined for extended systems as the corresponding charge distributions are neutral.
Since the lowest non-zero multipole of $\Delta_n$ is the quadrupole, $\mathcal E_2$ in Eq.~\eqref{eqn:error-e2} is (at worst) a sum of quadrupole-quadrupole interactions, decaying as $r^{-5}$. 
Only the part of that sum involving terms with (roughly) $r>r_c$ cannot be learned by the DP model. 
These terms contribute to the error. Taking 6~\AA\  for a typical $r_c$ value, the sum would include terms smaller than 1~meV.
Due to the fast rate of decay the double sum should converge rapidly. 
We could estimate $\mathcal E_1$ in Eq.~\eqref{eqn:error-e1} in a similar fashion by writing $G_t$ as a sum of local dipole contributions, resulting in a sum of dipole-quadrupole interactions that decay as $r^{-4}$, i.e., still rapidly but less rapidly than the terms in $\mathcal E_2$.
This estimate is, however, affected by the arbitrariness of the local dipole definition. 
A better way of estimating $\mathcal E_1$ is the following. Let $E_\Delta(r)$ be the electric field generated by the sum of the $\Delta_n$ distributions:
\begin{align}\label{eqn:error-efield}
    E_\Delta(r) = \int dr' \frac{r - r'}{\vert r - r'\vert^3} \sum_n\Delta_n(r)
\end{align}
To leading order, the electric field in Eq.~\eqref{eqn:error-efield} is a sum of quadrupole contributions that decay like $r^{-4}$ for $r > r_c$. This field is essentially uniform on the molecular scale. Let us call the corresponding average field $\bar E_\Delta$. Then, the average energy error can be written in terms of the cell dipole (polarization times volume) $M$ as:
\begin{align}\label{eqn:error-e1-me}
    \mathcal E_1 = - M \cdot \bar E_\Delta
\end{align}
In extended systems $M$ is defined modulo a quantum\recheck{\cite{vanderbilt2018berry}}. Thus, the error in Eq. \eqref{eqn:error-e1-me} is only well defined for a change of ionic coordinates that leads to a change of $M$. 

In summary, the error in the LR contribution to the electrostatic energy, when the true electronic distribution $\rho_e$ is replaced by 
the Gaussian distribution $G_e$, is given by a sum of small terms that decay at most as $r^{-4}$ for $r > r_c$. 
It would be possible to further reduce these errors by representing $G_e$ as a sum of non-spherical Gaussian distributions
having the spreads of the Wannier centroid distributions, which could be learned by an extended DW approach. In this way 
the lowest non-zero multipole of $\Delta_n$ would be the octupole. At $r_c$=6~\AA, such treatment would reduce the error 
by approximately an order of magnitude. However, this formulation would also result in a significant increase of the computational
complexity, which was deemed unnecessary based on our tests.

We remark that formulations of LR electrostatics based on effective atomic charges cannot reproduce the fluctuations of $M$ \recheck{in condensed phase}. In these formulations, the atom centered 
electronic charge distribution differ locally from the true distribution $\rho_e$ at the dipole rather than at the quadrupole 
level. 
Therefore, 
%\recheck{by using Eq.~\eqref{eqn:error-e1} and letting the lowest non-zero dipole of $\Delta_n$ being dipole}, 
the corresponding error would be associated with dipole-dipole interactions that are only conditionally convergent. 
\recheck{It is worth noting that, in finite systems, the error is reduced by adopting partial charge models trained to reproduce the molecular dipole.}

\subsection{Deep Potential Long-Range model} 

In DPLR the PES is given by
\begin{align}\label{eqn:e-decomp}
    E = E_\dpsr + E_{G_t},
\end{align}
where the electrostatic energy $E_{G_t}$ was introduced in Eq.~\eqref{eqn:e-dplr}, %\MCM{$E_{lr}$ in fact contains both short-range and long-range electrostatic, correct? Perhaps it would make more sense to term it as $E_{Coul}$? In the way that is written, one could think that $E_{lr}$ contains only the long-range part of electrostatic interactions. I agree and have changed the text accordingly} 
and $E_\dpsr$ is a short-range
contribution constructed as in the standard DP model (see Appendix~\ref{sec:app-srdp} for more details). However, $E_\dpsr$
is not equal to the standard DP energy, because the Gaussian electrostatic energy $E_{G_t}$ is subtracted from the DFT training data to avoid double counting.  

The forces and the virial are derivatives of the energy with respect to the atomic positions and the simulation cell tensor, respectively.
Within DPLR, the energy $E$ depends on the atomic coordinates, both explicitly and implicitly, via the dependence of the WCs on the atomic coordinates. The implicit dependence makes the calculations nontrivial.

The force on atom $I$ is given by
\begin{align}\label{eq:tmp-a1}
    F_I = 
    - \frac{\partial E}{\partial R_I} 
    =
    - \frac{\partial E_\dpsr}{\partial R_I} 
    - \frac{\partial E_{G_t}}{\partial R_I} 
    - \sum_n \frac{\partial E_{G_t}}{\partial W_n} 
    \frac{\partial W_n}{\partial R_I}. 
\end{align}
Here the environmental dependence of $W_n$ is given by the DW model. We recall that there is a one-to-one correspondence 
between the Wannier centroid $n$ and its nearest atom, the atom $I$. This correspondence establishes a bijective mapping between the indices of the WCs and the indices of a subset of the ions, that we indicate, equivalently, by $n = n(I)$ or by $I = I(n)$. Due to the nearsightedness of the electronic matter~\cite{}, the position of the WC associated with atom $I$ depends on the local environment
of that atom within a cutoff radius $r_c$ that can be safely assumed to be equal to the cutoff radius of the DP model. 
Indicating with $\cwcdisp_n$ the displacement of $W_n$ relative to $R_I$, we have: 
\begin{align}
    W_n = R_{I(n)} + \cwcdisp_n,
\end{align}
which relates the environmental dependence of $\cwcdisp_n$ to that of $W_n$. The latter 
is smooth by construction in the DW model.
Thus,
\begin{align}
    \frac{\partial W_n}{\partial R_J}
    = 
    \delta_{I(n),J} + \frac{\partial \cwcdisp_n}{\partial R_J},
\end{align}
where $\delta_{I,J}$ is the Kronecker delta, and the force on atom $I$ in Eq. \eqref{eq:tmp-a1} becomes
\begin{align}\label{eq:app-f}
     F_I = 
     - \frac{\partial E_\dpsr}{\partial R_I} 
    - \frac{\partial E_{G_t}}{\partial R_I} 
    - \frac{\partial E_{G_t}}{\partial W_{n(I)}} 
    - \sum_n \frac{\partial E_{G_t}}{\partial W_n} 
    \frac{\partial \cwcdisp_n}{\partial R_I}. 
\end{align}
The first term on the right-hand-side (RHS) of Eq. \eqref{eq:app-f} is the standard DP force, the second and third terms give 
the electrostatic force on atom $I$ and its associated WC, respectively, while
the fourth term originates from the environmental dependence of the WC.
The derivatives $\partial \cwcdisp_n/\partial R_I$ are calculated by back-propagating the DW model.

We indicate the cell tensor by $h = \{h_{\alpha\beta}\}$, with $h_{\alpha\beta}$ being the $\beta$-th component of $\alpha$-th cell vector. 
%\recheck{It is noted that the $\beta$ as a subscript is distinguished from the spread parameter.}
The virial is defined by
\begin{align}\label{eq:app-v0}
\Xi_{\alpha\beta} 
= -\frac{\partial E}{\partial h_{\gamma\alpha}} h_{\gamma\beta},
\end{align}
where summation over repeated indices is assumed.
By using Eq.~\eqref{eqn:e-decomp} we have
\begin{align}\label{eq:app-v1}
\Xi_{\alpha\beta} = -\frac{\partial E_\dpsr}{\partial h_{\gamma\alpha}} h_{\gamma\beta}  -\frac{\partial E_{G_t}}{\partial h_{\gamma\alpha}} h_{\gamma\beta}
= \Xi^\dpsr_{\alpha\beta} -\frac{\partial E_{G_t}}{\partial h_{\gamma\alpha}} h_{\gamma\beta},  
\end{align}
The first term on the RHS of Eq.~\eqref{eq:app-v1} is the short-range virial contribution, $\Xi^\dpsr_{\alpha\beta}$,
which is calculated as in the standard DP model. The remaining term is the electrostatic virial contribution, whose calculation is non-trivial as we detail below. 
% Firstly, remind that it is given by
% \begin{align}
%     E_\dplr = 
%     \frac1{2\pi V} \sum_{m\neq0}
%     f(m)S^2(m), \quad 
%     S(m) = \sum_I q_I e^{2\pi imR_I} +\sum_n q_n e^{2\pi imW_n}
% \end{align}
% where we have introduced the short-hand notation of 
% $f(m) = \frac1{m^2}{\exp(-\pi^2m^2/\beta)}$.
First, it is convenient to define the reciprocal cell tensor $h^{-1} = \{h^{-1}_{\beta\alpha}\}$, in which $h^{-1}_{\beta\alpha}$ is the $\beta$-th component of $\alpha$-th reciprocal cell vector. The direct and reciprocal cell tensors define linear scaling transformations in direct and
reciprocal space that are useful in MD simulations with variable cell.
In direct space, a scaled coordinate $s$ is related to its non-scaled counterpart $R$ by $R = sh$, and, in reciprocal space, 
the lattice vector $m$ is related to its scaled counterpart $k$ by $m = h^{-1}k$.
%\RC{Shall we use m = k h^{-1}? Eventually, it all boil down on how we define the scalar product}
%\WH{It is more intuitive to use h^{-1}k, because the second index of h^{-1} contracts with k.}
As a consequence, the scalar product 
$mR = sk$ is independent of the cell tensor $h$, and so is the factor $e^{2\pi imR_I}$ in Eq.~\eqref{eqn:sm}. However, the factor $e^{2\pi imW_I}$ in the same equation does depend on the cell vector $h$, because in general 
the positions of the WCs do not change linearly with a cell deformation.
%\RC{in Eq. 10 the centroids appear as W_n, not as W_I. Perhaps we should introduce here too the notation W_n(I) }
Then, we can write for the electrostatic contribution to the virial:
\begin{align}\label{eq:tmp-a4}
    -\frac{\partial E_{G_t}}{\partial h_{\gamma\alpha}} h_{\gamma\beta}
    =
    \Xi^{\rec}_{\alpha\beta}
    +
    \Xi^{c}_{\alpha\beta},
\end{align}
with
\begin{align}\label{eq:tmp-a5}
    \Xi^{\rec}_{\alpha\beta} & = 
    -\frac{\partial}{\partial h_{\gamma\alpha}} \Big(\frac1{2\pi V} \Big)
    h_{\gamma\beta} \sum_{m\neq0}f(m)S^2(m)
    -
    \frac1{2\pi V} \sum_{m\neq0}\frac{\partial f(m)}{\partial h_{\gamma\alpha}} h_{\gamma\beta} S^2(m) \\ \label{eq:tmp-a6}
    \Xi^{c}_{\alpha\beta} &= 
    -
    \frac1{\pi V} \sum_{m\neq0}f(m)
    S(m) \frac{\partial S(m)} {\partial h_{\gamma\alpha}} h_{\gamma\beta}, 
\end{align}
% \begin{align}\nonumber
%     -\frac{\partial E_\dplr}{\partial h_{\gamma\alpha}} h_{\gamma\beta}
%     = &
%     -\frac{\partial}{\partial h_{\gamma\alpha}} \Big(\frac1{2\pi V} \Big)
%     h_{\gamma\beta} \sum_{m\neq0}f(m)S^2(m)
%     -
%     \frac1{2\pi V} \sum_{m\neq0}\frac{\partial f(m)}{\partial h_{\gamma\alpha}} h_{\gamma\beta} S^2(m) \\ \label{eq:tmp-a6}
%     &-
%     \frac1{\pi V} \sum_{m\neq0}f(m)
%     S(m) \frac{\partial S(m)} {\partial h_{\gamma\alpha}} h_{\gamma\beta} \\ \label{eq:tmp-a7}
%     = & \,
%     \Xi^{\rec}_{\alpha\beta}
%     +
%     \Xi^{\corr}_{\alpha\beta}
% \end{align}
where we used the definition $f(m) = \frac1{m^2}{\exp(-\pi^2m^2/\beta)}$.
$\Xi^\rec_{\alpha\beta}$ in Eq.~\eqref{eq:tmp-a5} is the standard reciprocal virial contribution of the Ewald method, while the correction term in Eq.~\eqref{eq:tmp-a6} accounts for the nonlinear dependence of the WCs on the cell tensor. 
The correction term $ \Xi^{c}_{\alpha\beta}$ is calculated as follows
\begin{align}\label{eq:tmp-a8}
    \Xi^{c}_{\alpha\beta}
    =
    -
    \frac1{\pi V} \sum_{m\neq0}f(m) S(m)
    \sum_n q_n e^{2\pi iW_n m} 2\pi i 
    \Big(
    \frac{\partial W_{n\delta}}{\partial h_{\gamma\alpha}} m_\delta +
    W_{n\delta} \frac{\partial m_\delta}{\partial h_{\gamma\alpha}}
    \Big)
    h_{\gamma\beta}, 
\end{align}
where the terms in parentheses are given by
\begin{align*}
    \frac{\partial W_{n\delta}}{\partial h_{\gamma\alpha}} m_\delta h_{\gamma\beta}
    &= 
    \sum_J \frac{\partial W_{n\delta}}{\partial R_{J\epsilon}}\frac{\partial R_{J\epsilon}}{\partial h_{\gamma\alpha}} m_\delta h_{\gamma\beta}
    =
    \sum_J \frac{\partial W_{n\delta}}{\partial R_{J\alpha}} R_{J\beta} m_\delta
    = 
    R_{n\beta} m_\alpha
    +
    \sum_J \frac{\partial \cwcdisp_{n\delta}}{\partial R_{J\alpha}} R_{J\beta} m_\delta\\
    W_{n\delta} \frac{\partial m_\delta}{\partial h_{\gamma\alpha}} h_{\gamma\beta}
    & =
    W_{n\delta} \frac{\partial h^{-1}_{\delta\epsilon}}{\partial h_{\gamma\alpha}} k_\epsilon h_{\gamma\beta}  
    =
    -W_{n\delta}h^{-1}_{\delta\eta} \frac{\partial h_{\eta\zeta}}{\partial h_{\gamma\alpha}} h^{-1}_{\zeta\epsilon} k_\epsilon h_{\gamma\beta}
    =
    -W_{n\delta}h^{-1}_{\delta\gamma}  h^{-1}_{\alpha\epsilon} k_\epsilon h_{\gamma\beta}
    = - W_{n\beta} m_\alpha
\end{align*}
Thus, the correction virial term in Eq.~\eqref{eq:tmp-a8} becomes
\begin{align}
    \Xi^{c}_{\alpha\beta}
    =
    -
    \frac1{\pi V} \sum_{m\neq0}f(m) S(m)
    \sum_n q_n e^{2\pi iW_n m} 2\pi i 
    \Big(
    \cwcdisp_{n\beta}m_\alpha 
    +
    \sum_J \frac{\partial \cwcdisp_{n\delta}}{\partial R_{J\alpha}} R_{J\beta} m_\delta
    \Big)
\end{align}
Given that 
\begin{align}
    -\frac{\partial E_{G_t}}{\partial W_{n\alpha}}
    =
    -
    \frac1{\pi V} \sum_{m\neq0}f(m) S(m)
    2\pi i m_\alpha q_n e^{2\pi iW_n m}, 
\end{align}
the correction virial simplifies to
\begin{align}\label{eq:tmp-a11}
     \Xi^{c}_{\alpha\beta}
     =
     - \sum_n \frac{\partial E_{G_t}}{\partial W_{n\alpha}} \cwcdisp_{n\beta} 
     -
     \sum_{nJ} \frac{\partial E_{G_t}}{\partial W_{n}}
     \frac{\partial \cwcdisp_{n}}{\partial R_{J\alpha}} R_{J\beta}
\end{align}
Finally, using Eqs.~\eqref{eq:app-v1}, \eqref{eq:tmp-a4}, \eqref{eq:tmp-a5} and  \eqref{eq:tmp-a11}, the virial takes the more compact form
\begin{align}
    \Xi_{\alpha\beta} = \Xi^\dpsr_{\alpha\beta} + 
    \Xi^\rec_{\alpha\beta}
    - \sum_{nJ} \frac{\partial E_{G_t}}{\partial W_{n}}
     \frac{\partial \cwcdisp_{n}}{\partial R_{J\alpha}} R_{J\beta}
    - \sum_n \frac{\partial E_{G_t}}{\partial W_{n\alpha}} \cwcdisp_{n\beta} 
\end{align}

\begin{figure}
    \centering
    \includegraphics[width=0.45\textwidth]{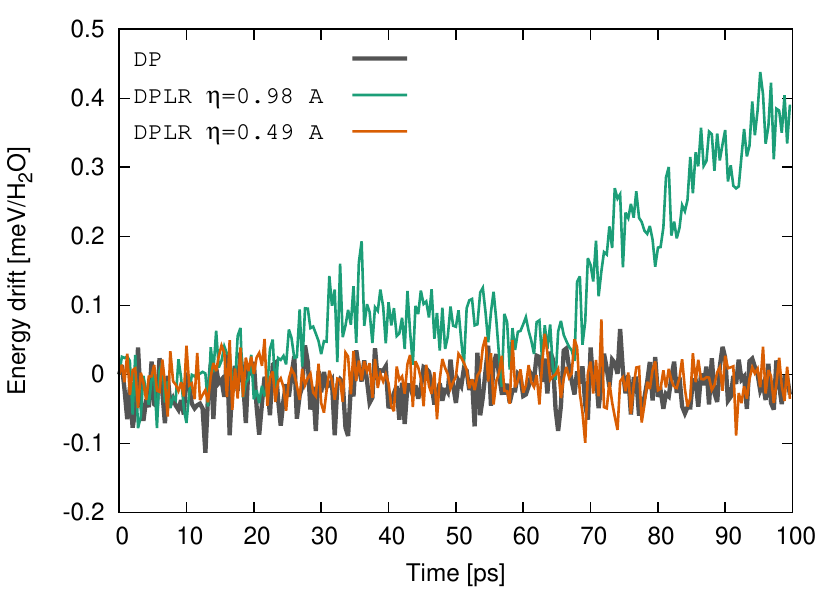}
    \caption{Drift of the total energy of DP and DPLR models in a 100~ps NVE simulation of 128 water molecules. 
    The time-step is set to 0.5~fs.
    Two mesh spacings of the PPPM method, i.e.~$\eta = 0.98$~\AA\ and $\eta = 0.49$~\AA, are used in the DPLR model.
    }
    \label{fig:ener_drift}
\end{figure}

We checked that the above formulae are correct by comparing the analytical derivatives in the formulae to the numerical    
derivatives calculated with finite differences. As a consequence,
in molecular dynamics simulations the conservation laws associated with the equations of motion should be satisfied within the accuracy of the adopted numerical implementation. 
We find, for instance, that in NVE trajectories of liquid water with 128 molecules, the total energy shows a small drift of approximately $0.4$~meV/\chem{H_2O} ($\sim$1~K) per 100 ps with the DPLR approach, when using a mesh spacing $\eta=0.98$~\AA\ in the calculation of the Ewald sum. As shown in Fig.~\ref{fig:ener_drift}, this is small but greater than the drift of a DP trajectory for the same system, which is almost not observable in a 100~ps trajectory.
The total energy drift in the DPLR trajectory is controlled by the numerical accuracy of the fast algorithm for computing the Ewald sum, here the PPPM method, and can be further reduced by using a finer mesh.
As shown in Fig.~\ref{fig:ener_drift}, the drift of DPLR is not observable when the mesh spacing $\eta$ is set equal to $0.49$~\AA.

% This is small but greater than the drift affecting the same system in DP trajectories.
% The reason for this difference between DP and DPLR is that the drift is controlled by the numerical accuracy of the Verlet algorithm in the DP case, while it is controlled by the numerical accuracy of the technique for computing Ewald sums, such as the particle-particle-particle-mesh method, in the DPLR case.    

% The force of the DPLR model is given by
% \begin{align}
%     F_I = 
%      - \frac{\partial E_\dpsr}{\partial R_I} 
%     - \frac{\partial E_\dplr}{\partial R_I} 
%     - \frac{\partial E_\dplr}{\partial W_I} 
%     - \sum_J \frac{\partial E_\dplr}{\partial W_J} 
%     \frac{\partial \Delta_J}{\partial R_I} 
% \end{align}
% and virial is derived as
% \begin{align}
%     \Xi_{\alpha\beta} = \Xi^\dpsr_{\alpha\beta} + 
%     \Xi^\rec_{\alpha\beta}
%     - \sum_{IJ} \frac{\partial E_\dplr}{\partial W_{I}}
%      \frac{\partial \Delta_{I}}{\partial R_{J\alpha}} R_{J\beta}
%     - \sum_I \frac{\partial E_\dplr}{\partial W_{I\alpha}} \Delta_{I\beta} 
% \end{align}
% The details on the derivation are provided in Appendix~\ref{sec:app-fv}.

\section{Deep Potential Long-Range model for water}

We construct a DPLR model for water based on the PBE functional approximation of DFT~\cite{perdew1996generalized}. 
% The choice of PBE as the reference DFT model is due to the relative simplicity of this functional approximation that facilitates the construction of the corresponding DP, DW, and DPLR models. 
It is well known that PBE substantially overestimates the strength of the hydrogen-bonds in water. 
As a consequence, it fails to describe correctly some important thermodynamic properties, like the relative density of ice and water at ambient pressure~\cite{gaiduk2015density}.
However, here we are not interested in constructing a state-of-the-art water model, but rather we intend to demonstrate that DPLR can capture long-range electrostatic effects missing in the DP model.
\recheck{For this purpose, the choice of the exchange-correlation functional is not a critical issue. It would be straightforward to adopt more accurate functionals than PBE by simply replacing PBE with any other functional of choice, in the labeling steps whereby energy, force and virial tensor are calculated at selected atomic configurations.
}
% % For that purpose PBE is good enough. 
% It would be straightforward to adopt more accurate functionals than PBE, such as, e.g., SCAN~\cite{sun2015strongly}, by simply replacing PBE with SCAN, or any other functional of choice, in the labeling steps whereby energy, force and virial tensor are calculated at selected atomic configurations. We note that the PBE functional, which has a relatively simple dependence on the density and its gradient, is numerically more stable than the SCAN functional, which depends on the kinetic energy density in addition to the density and its gradient~\cite{furness2019enhancing,bartok2019regularized}. 
% Thus, for the purpose of comparing DPLR and its DP counterpart, PBE is more convenient than SCAN. 
In the next three subsections, we construct DP, DW, and DPLR models based on the PBE functional approximation.

\subsection{Training data and DP model}

We use the concurrent learning scheme Deep Potential GENerator (DP-GEN)~\cite{zhang2019active,zhang2020dp}. DP-GEN iteratively enlarges the training dataset of labeled DFT data and refines a representative ensemble of DP models by learning from a growing dataset. In this work the representative ensemble consists of four models that differ in the random initialization of the network parameters but are otherwise trained on the same data. In the DP-GEN protocol 
the four DP models are used to sample the relevant thermodynamic space with deep potential molecular dynamics (DPMD) trajectories. The DP construction includes three stages.
In the first one (iteration 1 to 7), bulk configurations in the thermodynamics space $200 \leq T \leq 400$~K and $1\leq P \leq 10^4$~Bar are explored with isothermal-isobaric (NPT) DPMD. 
The initial configurations for bulk NPT DPMD simulations are snapshots of DPMD simulations with 128 molecules in the liquid state obtained with the DP model developed in Ref.~\citen{zhang2018deep}.
The second stage extends from iteration 8 to iteration 15. In this stage slab configurations are explored with canonical (NVT) DPMD simulations in the temperature range $200 \leq T \leq 400$~K.
The initial configurations are prepared by adding a vacuum region on top of bulk liquid configurations. 
Different surface structures with thickness of the vacuum region ranging from 0.5~\AA\ to  7.0~\AA\ are considered.
In the third stage, from iteration 16 to 23, low density bulk configurations are explored by NVT simulations for $200 \leq T \leq 400$~K. Initial configurations are created by randomly removing 64, 96, 112 and 120 molecules from the initial configurations of stage 1.
At each stage, the time span of the DPMD simulations is gradually increased over the iterations, from 1~ps to 10~ps.  
The maximal deviation of the forces predicted by the ensemble of DP models is monitored along the trajectories and used to classify the explored configurations. 
Configurations with deviation between 0.15 to 0.30~eV/\AA\ are labelled as training data, following the protocol detailed in Ref.~\citen{zhang2020dp}.
Single-shot DFT calculations are performed at these configurations.

DFT calculations are carried out with the Vienna {\it ab initio} simulation package (VASP) version 5.4.4~\cite{kresse1996efficiency,kresse1996efficient},
using the PBE exchange-correlation energy. 
The kinetic energy cut-off for the planewave basis is set to 1500~eV. 
The self-consistent field procedure is stopped when the energy difference between the last two iterations is less than $10^{-6}$~eV.  
% The VASP parameter \texttt{LASPH} is switched on to include the non-spherical gradient correction contributions in the PAW sphere. 

The DP models are trained with the DeePMD-kit package~\cite{wang2018deepmd}. 
The cut-off radius $r_c$ for the atomic environments is set to 6~\AA. 
The size of the embedding and that of the fitting network are set to (25, 50, 100) and (240, 240, 240), respectively. 
The DP models are trained with the Adam stochastic gradient descent scheme~\cite{kingma2015} using $10^6$ steps, during which the learning rate exponentially decays from $10^{-3}$ to $3.5\times 10^{-8}$.
At each training step, a subset of the training dataset, referred to as the mini-batch, is used to calculate the gradient of the loss function. 
The size of the minibatch is 1, i.e.~only 1 configuration is used for calculating the gradient.

At the end of each training stage the fraction of configurations with gradient deviation larger than 0.15~eV/\AA\ is reduced to less than 0.1\%, indicating satisfactory convergence of the DP-GEN cycle.
A total of 448 configurations are selected as training data during the DP-GEN protocol.
They include 323 bulk,  94 surface, and 31 low density configurations.
The dataset used for training the initial DP models has 135 configurations, thus the total training data includes 583 configurations. 

% By using the DP model at the last step of DP-GEN iteration, an NVT DPMD simulation of 200 ps is performed at 300~K for a slab configuration of water, which has 128 molecules and a 16.3\AA $\times$ 16.3\AA $\times$ 23.6\AA\ simulation region. The configurations are recorded every 2~ps along the DPMD trajectory, and then are labeled with energy, force, virial and CWC by DFT calculations. 89 out of the 100 configurations succeeded in DFT. They serve as an independent test dataset for validate the models. 

\subsection{DW model}

\begin{figure}
    \centering
    \includegraphics[width=0.7\textwidth]{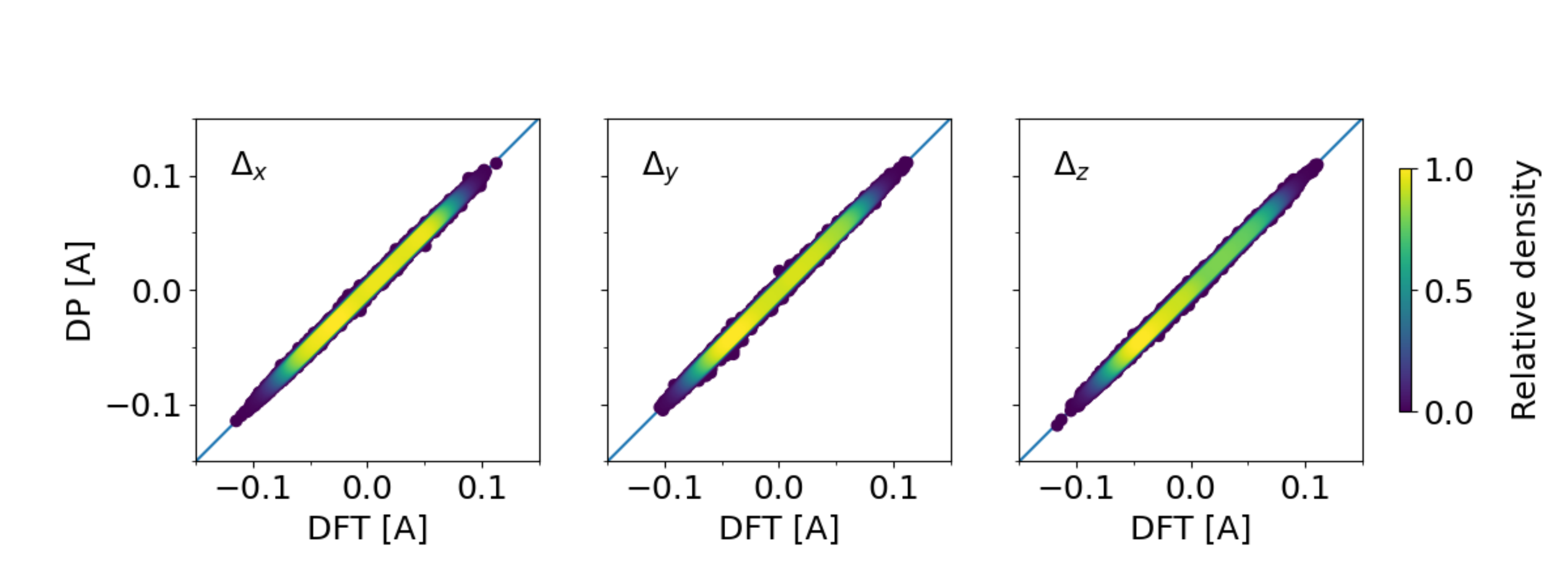}
    \caption{Training accuracy of the DW model for water. The three panels report the three Cartesian displacements of the Wannier centroids (WC) relative to their nearest oxygen atom predicted by DW (vertical axis) and by DFT (horizontal axis).}
    \label{fig:cwc-train-err}
\end{figure}

For each water configuration, the DW model gives the position of the WC associated with a given molecule relative to the O atom of that molecule. 
The training data are the same data used to generate the DP model.
Each WC is uniquely associated with a particular O atom and depends on the atomic environment of that atom within a cutoff radius, which is set here to 6~\AA, i.e., it is the same cutoff $r_c$ of the DP model. 
We use the standard DP descriptor $\mathcal D = \mathcal D^a$ (see Appendix for details).
The sizes of the embedding and fitting networks are (25, 50, 100) and (100, 100, 100), respectively. 
The model is trained with the Adam stochastic gradient descent method~\cite{kingma2015} using $10^6$ steps. 
The batch size is set equal to 1. The learning rate exponentially decays from $1.0 \times 10^{-2}$ to $5.6\times 10^{-8}$.
% The other hyper-parameters of the DW training are the same as those used for training the standard DP model. 

The training accuracy of the DW model is $1.7\times 10^{-3}$~\AA, which is much smaller than the typical distance $\cwcdisp$ of a WC from its reference O atom. 
The test accuracy is $1.9\times 10^{-3}$~\AA. 
Thus, the generalization gap between test and training errors is almost negligible.
The correlation between labelled and predicted $\cwcdisp$ is graphically illustrated in  Fig.~\ref{fig:cwc-train-err}.

\subsection{DPLR model}\label{sec:lrdp-train}

\begin{figure}
    \centering
    \includegraphics[width=0.5\textwidth]{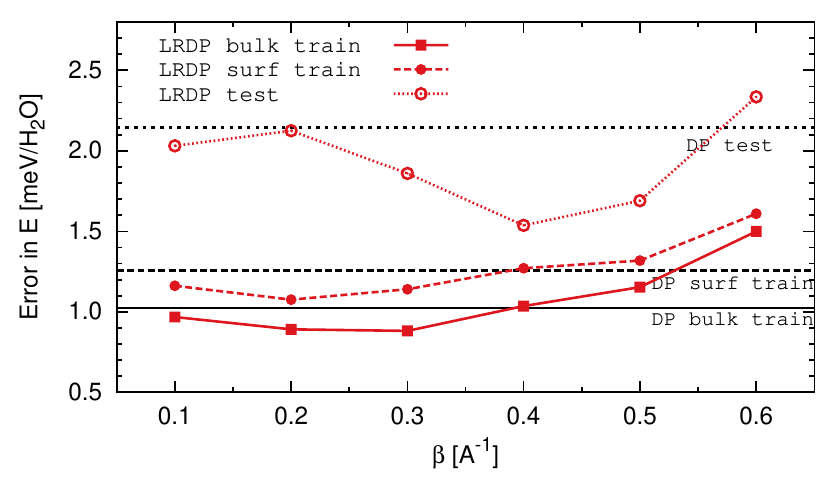}
    \caption{Root mean square error (RMSE) of the energy predicted by the DPLR model as a function of the spread parameter $\beta$. The training error of bulk (solid squares connected by a red solid line) and surface configurations (solid circles connected by a red dashed line), and the test error of surface configurations (open circles connected by a red dotted line) are shown (see text for more details).
    The corresponding errors for the DP model are plotted as a black solid line (bulk training error), a black dashed line (surface training error), and a black dotted line (test error).
    % Training and test errors of the standard DP model are shown as black lines for reference.
    }
    \label{fig:error}
\end{figure}

The DPLR model is trained on the dataset generated by DP-GEN.
In DPLR, the spread parameter $\beta$, i.e.~the inverse spread of the Gaussian charge distribution, needs to be fixed. 
In the limit of $\beta$ going to 0, the Gaussian width is infinite and DPLR reduces to the standard DP model.
On the other hand, for $\beta$ going to infinity, the Gaussian charges become point charges. 
%\AZP{The spread parameter $\beta$ is not purely a "free parameter", there should be some physical significance and in principle it could be obtained from the actual electronic wavefunctions. From Fig. 2, there is a broad range of $\beta$ values that give comparable results. I think it would be much better to justify the choice of $\beta$  (0.4 A-1) physically. }
In this limit, the magnitude of the intramolecular Coulomb force between the oxygen ion ($q_I=6e$) and its associated WC ($q_n=-8e$), for a typical separation distance ($\cwcdisp\approx 0.07$\AA), is about $10^4$~eV/\AA\ , which is about four orders of magnitude larger than the average atomic force. 
This situation would create serious numerical problems due to the difficulty of resolving a labeled force from the strong intramolecular Coulomb force.
Therefore, there should be an optimal $\beta$, neither too small to avoid reduction to the standard DP model, nor too large to avoid distributions too close to point-charges. 

To study the effect of the spread parameter on the accuracy of the DPLR models, we train different DPLR models with
% same short range cut-off of 6~\AA\ but  
different spread parameters ($\beta = \{0.1, 0.2,\dots,0.6\}$\AA$^{-1}$). 
To ensure proper convergence of the Ewald sum in Eq.~\eqref{eqn:e-dplr}, the mesh spacing $\eta$ is set equal to the largest value compatible with the constraint $\eta\cdot\beta \leq 0.4$.  
% For comparison, we also train a standard DP model with cut-off 6~\AA.
In DP and in the short-range part of the DPLR models we adopt a hybrid descriptor (see Appendix for details) $\mathcal D = (\mathcal D^a, \mathcal D^r)$, and the cut-off of $\mathcal D^a$ is set to $r_c^a = 4$~\AA, while the cut-off of $\mathcal D^r$ is set to $r_c = 6$~\AA. The size of the embedding net of  $\mathcal D^a$ is (25, 50, 100), and that of  $\mathcal D^r$ is (10, 20, 40).
Training and test errors, as a function of $\beta$, are shown in Fig.~\ref{fig:error}.
The training dataset is divided into two parts, one consisting only of bulk configurations and the other consisting only of surface configurations. 
The test dataset includes surface configurations with 128 molecules with a vacuum region of 7 or 12~\AA\ , generated along a 100-ps DPLR molecular dynamics  
(DPLRMD) trajectory for each of the two widths of the vacuum region. Configurations are recorded every 10~ps, and are labeled with \emph{ab initio} DFT calculations. Thus, the test dataset includes 20 surface configurations in total. 

In Fig.~\ref{fig:error}, one can see that the energy training error for the bulk configurations is smaller than the corresponding error for the surface configurations in both DP and DPLR models, reflecting 
the more difficult task of interpolating the PES for surface than for bulk configurations, due to the larger variance of the local environments in the former case. For $\beta < 0.5$~\AA$^{-1}$, the DPLR training errors are close to those of DP, but for $\beta \geq 0.5$~\AA$^{-1}$ they gradually
increase and become notably larger than the corresponding DP errors, a behavior that we attribute to numerical difficulties with
Gaussian charges that are too localized.
% In the case of bulk configurations the training error of DPLR model is marginally smaller than that of the standard DP model in range $\beta < 0.5$~\AA$^{-1}$, while in the case of surface configurations, the training error of DPLR is almost identical to that of the DP model in the same parameter range.

The test dataset is composed of independent configurations not included in the training dataset, and serves to assess the capability of the
the models to deal with unforeseen situations. 
In the case of the standard DP model, the test error (2.0~meV/\chem{H_2O}) is larger than the surface training error (1.2~meV/\chem{H_2O}). 
The gap, usually called generalization gap, between test and training errors indicates that a model overfits the training data and looses accuracy when generalized to cases not included in the training dataset. 
The DPLR model with a small $\beta$ has a generalization gap comparable to that of the standard DP model, as expected because for $\beta \rightarrow 0$~\AA$^{-1}$ DPLR reduces to DP.
However, for $\beta \approx 0.4$~\AA$^{-1}$, the generalization gap is minimal. In this case 
the training error and the test error are 1.3 and 1.5~meV/\chem{H_2O}, respectively,
indicating that inclusion of long-range electrostatics contributes to the model's generalization ability.
For $\beta \geq 0.5$~\AA$^{-1}$ the test error increases with $\beta$, as expected from numerical difficulties 
with charge distributions that are too strongly localized.

% Moreover, the energy training error of the standard DP model in the bulk configurations is 1.3~meV/\chem{H_2O}, and is consistent with the training accuracy of the DPLR model for $\beta \leq 0.4$~\AA$^{-1}$.
% This indicate that the  accuracy of the standard DP model in bulk systems is the same as the DPLR model, so the long-range effect can be safely neglected. 
% The energy training and test accuracy of the surface configurations are almost identical for the DPLR, which means that the model is well trained and no over-fitting is observable. 
% For small $\beta$, the energy test error of the DPLR is in good agreement with the standard DP model. 
% As $\beta$ increases, the test error firstly decrease, reaching a minimum between $0.4\sim 0.5$\AA$^{-1}$, then rapidly increases. 
% Considering the energy error in bulk configurations, the optimal $\beta$ is chosen to be 0.4\AA$^{-1}$.
% The force errors in the bulk and surface configurations are almost identical.
% The training and test force errors are also the same, which indicates no over-fitting exists. 
% The force errors presents a similar dependency on $\beta$ as the energy error: The error firstly decreases, reaches a minimum at $\beta = 0.4$\AA$^{-1}$, and then increases as the $\beta$ increases. 

Based on the above analysis, the optimal spread parameter $\beta$ should have a value of about 0.4~\AA$^{-1}$. It is interesting to note that, with this value of the spread parameter, the radial extent of the Gaussian distribution $g_n(r)$ about the oxygen atom is close to the physical extent of the (valence) electron density of a water molecule, as measured by the radius of the surface on which the electron density is approximately one order of magnitude smaller than its maximum value. Thus, with the optimal $\beta$ the spherical Gaussian distribution $g_n(r)$ approximates (roughly) the physical WC distribution $\rho_n(r)$ and the error $\Delta_n(r)$ in Eq.~\eqref{eqn:gn-dn} is minimized.  

In the rest of the manuscript, we set $\beta = 0.4$~\AA${}^{-1}$ , unless stated otherwise.
We note that simply increasing the cut-off radius of the standard DP model to 8~\AA\ only marginally reduces training and test errors. Indeed, when using $r_c=8$~\AA\ in place of $r_c=6$~\AA,  
the bulk and surface training errors are reduced from 1.02~meV/\chem{H_2O} and 1.26~meV/\chem{H_2O} to 1.01~meV/\chem{H_2O} and 1.18~meV/\chem{H_2O}, respectively, while the
test error is reduced from 2.14~meV/\chem{H_2O} to 2.08~meV/\chem{H_2O}.
This indicates that the long-range effect is non-trivial and can only be captured by including explicitly 
the long-range electrostatic interaction.
% It is noted that increasing the cut-off radius of the standard DP model to 6~\AA\ will reduce the test error to 2.0~meV/\chem{H_2O}. 
% Further enlarge the cut-off radius to 8~\AA\ will reduce the surface training and test errors of the DPLR model to 1.0 and 1.3~meV/\chem{H_2O}, respectively.
% Therefore, the long-range effect is non-trivial and could not be easily captured by a standard DP model with a large cut-off radius.

% When using this splitting parameter, the test accuracy of the DPLR model is slightly higher than the standard DP model. The energy accuracy increases from 1.65~meV/\chem{H_2O} to  1.42~meV/\chem{H_2O}, while the force accuracy increases from 59~meV/\AA\ to 50~meV/\AA. 
% It worth noting that when using a  cut-off radius larger than~$6$\AA, the discrepancy between the DPLR and the standard DP model vanishes, which indicates that the long-range effect can be effectively captured by the standard DP model with a reasonably large cut-off radius. 
% A non-trivial test case, in which the long-range effect plays an essential role, will be presented in the next section. 

\section{Application to two water systems}

We compare the performance of DPLR and DP in two test cases. In one we compute the potential energy of interaction of two water molecules 
as a function of the relative distance. 
In the other we compute the free energy profile a water molecule versus its distance from a liquid water slab. 
The second example requires the calculation of a thermal property of a relatively complex system. 
In both cases, long-range electrostatic interactions
among the electric dipoles of the water molecules play a role. 
These interactions, absent in the DP model, are described with sufficient accuracy by DPLR.

\subsection{Water dimer}
\begin{figure}
    \centering
    \includegraphics[width=0.45\textwidth]{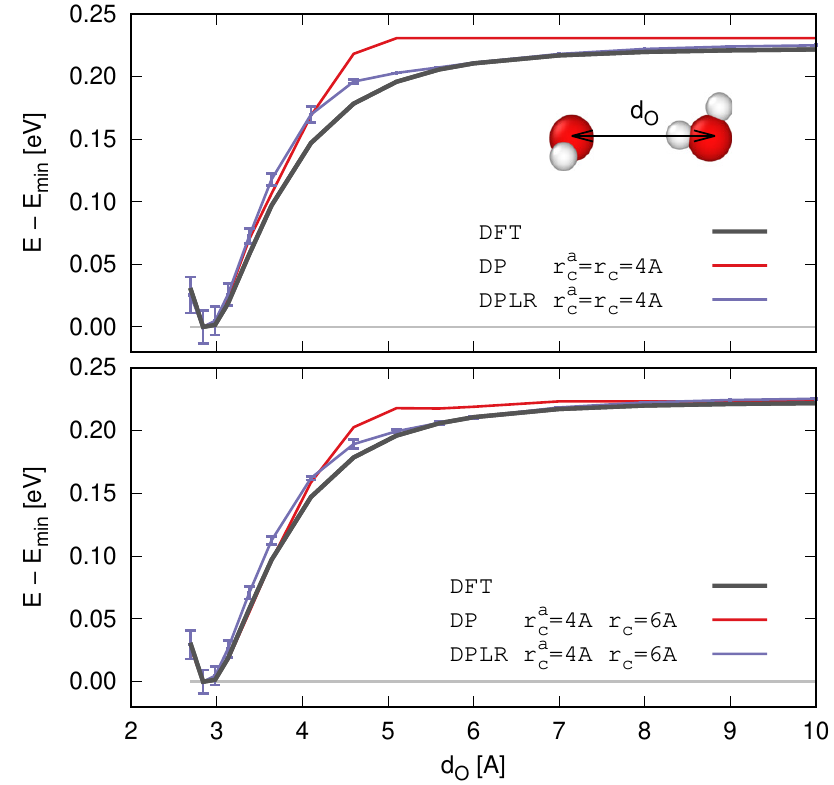}
    \caption{Potential energy of the water dimer as a function of the separation distance between the two water molecules. 
    The standard deviation of four independently trained DPLR models with the same spread parameter $\beta = 0.4$~\AA$^{-1}$ is shown by the error bars in the plot. The top panel compares DFT with DP and DPLR models with cutoff radii $r_c^a = r_c=4$\AA.
    \recheck{The bottom panel compares DFT with DP and DPLR models with cutoff radii $r_c^a = 4$~\AA\ and $r_c=6$\AA .
    All energies are referred to the potential energy minimum $E_{min}$.}
    % , which is reproduced accurately by all the DNN models
    }
    \label{fig:dimer}
\end{figure}

The potential energy curve of a water dimer as a function of the separation distance $d_O$ between the two molecules is reported in Fig.~\ref{fig:dimer}. 
\recheck{Both DP and DPLR models describe equally well the potential energy at short distance.
The DP model does not correctly predict the energy curve at a distance beyond the cut-off radius. 
Due to the finite range of the DP model, the corresponding potential energy curve misses the
$1/d_O^3$ tail of the DFT reference, which is due to dipole-dipole interactions. 
% It is noted that error at  does not decrease as a larger (6 v.s.~4~\AA) cut-off radius is used. 
% Interestingly, the error of DP at distance $d_O = 10$~\AA\ is larger in the case of $r_c = 6$~\AA\ than that in the case of $r_c = 4$~\AA.
By contrast, the long-range Coulomb tail is recovered accurately by the DPLR model, as shown in both panels of Fig.~\ref{fig:dimer}.
The accuracy at intermediate distance improves by increasing the cutoff in all the models, as expected. 
Although DPLR has a higher ability of representing the energy surface, it is marginally less accurate than the DP model between 3.5 and 4.0~\AA. This can be explained by the fact that dimer configurations were not included in the training dataset.
From Fig.~\ref{fig:dimer} one concludes that DPLR is a superior model with both cutoffs.
}

\subsection{Free energy profile of a molecule at varying distance from a liquid slab}

% \begin{figure}
%     \centering
%     \includegraphics[width=0.45\textwidth]{fig-000-1.pdf}
%     \caption{The schematic plot of the water absorption on a slab.}
%     \label{fig:water-slab}
% \end{figure}

\begin{figure}
    \centering
    \includegraphics{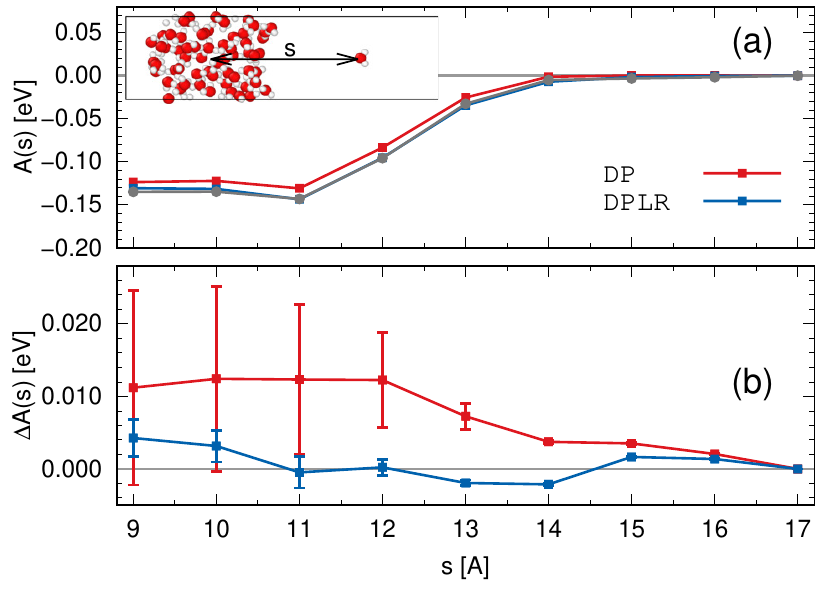}
    \caption{The free energy of water molecule absorption to a water slab.
    The panel (a) presents the free energy calculated by DFT, DPLR and DP models. 
    The panel (b) presents the error in free energy of the DPLR and DP models with respect to DFT.
    The error bar in the plot is given by the standard deviation of four independently trained models.
    The insert is a schematic plot of the system configuration.
    }
    \label{fig:fe}
\end{figure}

In this subsection, we test the performance of DPLR in a calculation of the
free energy profile of a water molecule as a function of its distance from a liquid water slab. The model has cut-off radii of $r_c^a = 4$~\AA\ and $r_c = 6$~\AA, spread parameter of 0.4~\AA$^{-1}$, and Ewald mesh spacing of 1~\AA. We simulate the slab and the adjacent vacuum region on a periodic box of size 50\AA$\times$ 11\AA $\times$ 11\AA. The slab contains 64 molecules, let to equilibrate with a long canonical MD run at a temperature of 300~K starting from an initial bulk liquid configuration. After equilibration, the half width of the slab is $\sim 7.9$~\AA ~and the adjacent vacuum region has a width of $\sim 34.2$~\AA.
We insert an additional molecule to this system, initially near the center of the vacuum region where it interacts weakly with the slab. Then, we vary the distance of the molecule from the slab and we calculate the free energy profile as a function of the distance, while keeping the whole system in equilibrium at 300~K. The distance of the molecule from the slab is defined by the distance of the respective centers of mass. A typical configuration is illustrated in the inset of Fig.~\ref{fig:fe} for a distance $s$ between the molecule and the liquid slab. In the atomic configuration $\{R\}$, the distance of the molecule from the slab is represented by the collective variable $\sigma(\{R\})$. In MD simulations an holonomic constraint, such as $\sigma(\{R\}) = s$, is easily imposed by a Lagrange multiplier. We set the initial distance to $s_{\textrm{far}} = 17$~\AA. At this distance the molecule experiences weak long-range dipolar interactions with the slab. We indicate by $F$ the projection of the force on the molecule along the direction connecting the center of mass of the molecule and that of the slab. When the molecule is displaced from $s_{\textrm{far}}$ to $s$, the corresponding free energy change, $A(s)$, is given by: 
\begin{align}\label{eq:pomf}
    A(s) = \int_s^{s_{\textrm{far}}} \langle F \rangle^\tau d \tau,
    \quad
    \langle F \rangle^\tau = \frac1Z \int F e^{-\beta H(\{R\})} \delta(\sigma(\{R\}) - \tau) d\{R\}.
\end{align}
Here $H(\{R\})$ and $Z$ are the Hamiltonian and the partition function of the system. 
For each value of $\tau$ the ensemble average in Eq. \eqref{eq:pomf} is calculated with canonical MD at 300~K for configurations that satisfy the constraint $\sigma(\{R\}) = \tau$. We compute the ensemble average at a discrete set of $\tau$ values using 1000~ps long MD trajectories, the first 200 ps of which serve for equilibration and are not included in the observation time. During observation configurations are recorded every 10~ps.
By varying $s$ from $s_{\textrm{far}}$ to $s_{\textrm{near}} = 9$~\AA, we obtain the blue free energy profile reported in panel (a) of Fig.~\ref{fig:fe} for the DPLR model. We divide the interval $[s_{\textrm{near}}, s_{\textrm{far}}]$ into 8 equal intervals and calculate the corresponding free energy changes by
evaluating the integrals in Eq. \eqref{eq:pomf} numerically with the trapezoidal rule.   
%We calculate the absorption free energy $A(s)$ by numerically integrating the mean force $\langle F \rangle^\tau$, i.e.
% We choose the distance, denoted by $s$, between the center of mass (COM) of the water slab and the COM of the water molecule as the reaction coordinate, and calculate the absorption free energy $A(s)$ by numerically integrating the mean force, i.e.
%\begin{align}\label{eq:pomf}
%    A(s) = \int_s^{\infty} \langle F \rangle^\tau d \tau,
%    \quad
%    \langle F \rangle^\tau = \frac1Z \int F e^{-\beta H(\{R\})} \delta(\sigma(\{R\}) - %\tau) d\{R\},
%\end{align}
%where $H(\{R\})$ and $Z$ indicate the Hamiltonian and the partition function, %respectively. 
For $s$ in the vicinity of $s_{\textrm{near}}$ the tagged molecule forms hydrogen bonds with neighboring molecules in the slab, suggesting that incorporation into the slab is taking place.

We use a simple approximation to estimate the deviation of the calculated DPLR profile from the DFT reference. Instead of re-weighting with the DFT energy the configurations along DPLR trajectories, which would be expensive, given the large number of configurations needed for statistical accuracy, we keep the DPLR weights and approximate the ensemble average with the following expression, which only requires DFT calculations at a relatively small number of configurations:
\begin{align}\label{eq:approx}
    \langle F_\dft \rangle^\tau_\dft \approx \langle F_\dft \rangle^\tau_\dplrc.
\end{align}
This procedure can be justified because DPLR reproduces accurately DFT energies with an average error of $\sim 1$~meV/\chem{H_2O}. 
We find that 80 configurations, equally spaced in time along an equilibrated 800~ps long trajectory, are sufficient to compute the average force at each $\tau$ value. 
The results are reported in panel (b) of Fig.~\ref{fig:fe}, where the baseline DFT reference corresponds to $\Delta A(s) = 0$. 
The deviations of DPLR from the baseline, calculated at discrete distances in the interval $[s_{\textrm{near}}, s_{\textrm{far}}]$, appear in the same figure as blue points connected by blue lines. 
The blue points represent the average of four independent DPLR models trained on the same data with same hyper-parameters but different random initialization of the model parameters.
The blue error bars are the corresponding standard deviations of the predictions of the four models. Overall, the deviation of DPLR from DFT is quite small and the uncertainty of the model, estimated from the standard deviation of the four models, is even smaller.     

We now consider a DP model with same cutoff radius of DPLR but without explicit long-range Coulomb interactions. To estimate the deviation of this model from DPLR  we adopt a similar approximation to the one adopted in Eq.~\eqref{eq:approx} , i.e., we assume:
\begin{align}
    \langle F_{\textrm{DP}} \rangle^\tau_{\textrm{DP}} \approx \langle F_{\textrm{DP}}\rangle^\tau_\dplrc,
\end{align}
and use the same configurations as before to calculate the ensemble averages and the free energy differences. The deviations of DP from DPLR and from the baseline DFT reference are reported as red points with error bars connected by red lines in panel (b) of Fig.~\ref{fig:fe}. As before, the DP points are averages of four independent models trained on the same data with different random initialization. The average free energy profile obtained in this way is reported as a red line in panel (a) of the same figure, showing that when the molecule approaches the slab the gain in free energy is smaller for DP than for DPLR, a result consistent with the missing attractive dipolar interactions in the DP model. However, the most important outcome of this analysis is that the DP model is affected by a large uncertainty. When evaluated along DPLR trajectories the standard deviation of the independent DP models grows rapidly as the tagged molecule approaches the slab to become almost an order of magnitude larger than the uncertainty of the corresponding DPLR model when the tagged molecule is near the slab. We attribute this behavior to the absence of long-range electrostatic interactions in the DP model as opposed to DPLR and DFT.        

\section{Phonons in crystalline sodium chloride}

\begin{figure}
    \centering
    \includegraphics[width=0.8\textwidth]{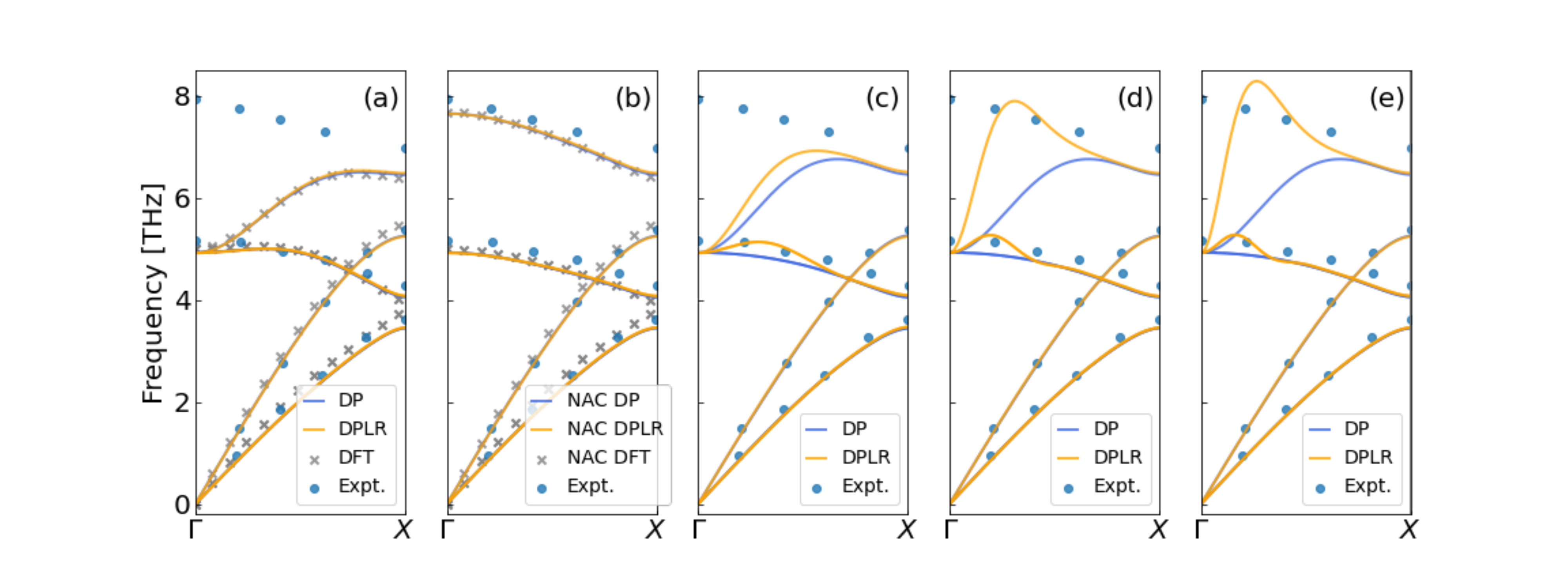}
    \caption{The LO-TO splitting of the phonon spectra of NaCl crystal.
    DPLR, DP, DFT and experiment (Expt.)~\cite{raunio1969phonon} are compared. 
    (a) Spectra obtained from the analytical dynamical matrix (see text) on a $2\times2\times2$ (64 atoms) supercell.
    (b) Spectra obtained from the full dynamical matrix, including  analytical and non-analytical correction (NAC) terms (see text), on a $2\times2\times2$ supercell.
    (c-e) Spectra obtained from the analytical dynamical matrix on a $3\times3\times3$ (216 atoms) (c), $5\times5\times5$ (1000 atoms) (d), and $6\times6\times6$ (1728 atoms) (e) supercell, respectively.
    }
    \label{fig:nacl}
\end{figure}

In this section we consider the phonon dispersion curves in the sodium-chloride (NaCl) crystal, as a further example to illustrate the importance of long-range electrostatic interactions. 
For the reference DFT model we adopt the PBE exchange-correlation functional and use pseudopotentials for the electron-ion interactions. DP and DPLR models for NaCl are trained on DFT data generated in a 200~ps DPMD NPT trajectory of a $2\times 2\times 2$ supercell (64 atoms) at 100~K and 1~Bar. In the DFT data we use  Brillouin Zone (BZ) sampling with  
a $4\times 4\times 4 $ Monkhorst Pack mesh to calculate the interatomic potential and the maximally localized Wannier distributions. The hyper-parameters of the DP and DPLR models are the same as in the water models discussed earlier in the manuscript. Using the same arguments of 
Sect.~\ref{sec:lrdp-train}, we find that the optimal value of the spread parameter for NaCl is $\beta = 0.2$~\AA$^{-1}$. The smaller $\beta$ of NaCl compared to water reflects the fact that the electron density distribution about the Cl atom in NaCl is more delocalized than the one about the O atom in the water molecule. The training accuracy of the DP model is $7.6\times 10^{-4}$~eV/atom for energy and $2.0\times 10^{-3}$~eV/\AA\ for atomic forces,  similar to that of  the DPLR model, which is $7.7\times 10^{-4}$~eV/atom for energy and $2.6\times 10^{-3}$~eV/\AA\ for atomic forces.

To compute the phonons, we calculate numerically the force constants from the second derivatives of the PES with respect to atomic displacements. We then obtain the phonon frequencies at a generic point $\bm q$ of the BZ of the crystal with the \texttt{phonopy} package~\cite{phonopy}, by diagonalizing the dynamical matrix given by a Fourier expansion of the force constants at wavevector $\bm q$. This approach would be adequate in non-polar materials where the force constants have finite range. In polar materials, like NaCl, long-range electrostatic interactions yield a non-analytic contribution to the dynamical matrix, which should be added to the analytic contribution calculated with the above procedure. We report in panel (a) of Fig.~\ref{fig:nacl} the phonon dispersion curves along the $\Gamma$-$X$ segment of the BZ obtained from the analytical part of the dynamical matrix by using a $2\times 2\times 2$ supercell for the force constants. DP and DPLR curves coincide within the accuracy of the calculation, and are very close to the corresponding DFT curves, as one could have expected given that a $2\times 2\times 2$ supercell was used for training the models. Experimental frequencies are also reported in the same panel, from which we see that, while the calculated acoustic and transverse optical (TO) modes agree with experiment, the calculated longitudinal optical (LO) modes deviate considerably from experiment, particularly in the long wavelength limit, where the calculated modes fail to exhibit an LO-TO splitting. Adding the non-analytic contribution to the dynamical matrix restores agreement between theory and experiment, as illustrated in panel (b) of Fig.~\ref{fig:nacl}. The non-analytic contribution to the dynamical matrix is given by~\cite{baroni2001phonons}:
\begin{align}\label{eqn:nac}
\begin{aligned}
     \tilde{D}^{\textrm{NA}}_{\kappa\alpha,\kappa'\beta}(\bm q)
    &=
    \frac{4\pi}{\Omega \sqrt{M_\kappa M_{\kappa'}}}
    \frac{ 
    (\sum_\gamma q_\gamma Z^\ast_{\kappa,\gamma\alpha})
    (\sum_{\gamma'} q_{\gamma'} Z^\ast_{\kappa',\gamma'\beta})
    }{
    \sum_{\alpha\beta} q_\alpha \epsilon_{\alpha\beta}^\infty q_\beta
    }
\end{aligned}
\end{align}
Here, $\Omega$ is the volume, $\kappa,\kappa'$ are atom indices, $\alpha,\beta,\gamma$ are Cartesian directions, ${M_\kappa},{M_{\kappa'}}$ are atomic masses, $Z^\ast_\kappa,Z^\ast_{\kappa'}$ are Born dynamical charge tensors, and $\epsilon^\infty$ is the dielectric tensor that describes the static response of the electrons at fixed ions. We recall that the dynamical Born charges are the derivatives of the polarization with respect to atomic displacements, and are therefore directly accessible from the DW network within DPLR. The dielectric tensor is related to the dielectric susceptibility or polarizability tensor $\chi$ via $\epsilon^\infty = 1+4\pi\chi$. The susceptibility is defined by the derivatives of the polarization with respect to an applied electric field and is accessible within a DW extension introduced to model the polarizability surface~\cite{sommers2020raman}. Thus, all the quantities needed for calculating the phonons in polar materials are accessible within the extended DP methodology. 

It is interesting to monitor the evolution of the analytical phonon modes calculated within DP and DPLR when the size of the supercell used to compute the force constants is increased. This is shown in panel (c), (d), and (e) of Fig.~\ref{fig:nacl} for a $3\times3\times3$ supercell (216 atoms), a $5\times5\times5$ supercell (1000 atoms), and a $6\times6\times6$ supercell (1728 atoms), respectively. The DP modes are essentially independent of size, indicating that, with the chosen cutoff radius ($r_c=6$~\AA), the force constants for atomic distances larger than those probed in a $2\times2\times2$ supercell are negligible. By contrast, the LO mode of DPLR is strongly size dependent, reflecting the appearance of not negligible force constants on larger supercells due to long-range dipolar interactions. With larger supercells the analytical spectra from DPLR recover a larger portion of the correct LO modes. However, the convergence is slow as the non-analytic behavior for $\bm q \rightarrow 0$ cannot be recovered from numerical Fourier interpolation. At small wavevectors we observe an upward bump in the LO and TO modes that becomes sharper and moves to smaller $\bm q$ values as the size of the supercell increases. The bump is a manifestation of the Gibbs oscillations that are expected to occur in place of the LO-TO splitting when attempting to reproduce the non-analytic behavior of the dynamical matrix with a Fourier sum~\cite{gonze1994interatomic}.

The long range contributions that are responsible for the spectral changes in panels (c), (d), and (e) derive from the Ewald contribution to the PES within DPLR. This example illustrates an essential advantage of DPLR over DP in materials where long-range electrostatic effects are important. While the DP model can describe well such materials on the supercell used for training, it is unable to extrapolate the PES to larger supercells, a drawback that is absent in the DPLR model.   
%\AZP{I do not quite see that the DPLR model completely eliminates the extrapolation error, based on the panels (c), (d) and (e). While better than DP, DPLR still has significant differences between experimental and predicted phonon curves. }

\section{External fields}

In this manuscript we have considered the electrostatic fields originating from internal charges, but, in the linear regime, it is straightforward to include the effect of an external time dependent field $\mathcal E_\textrm{ext}(t)$ that couples with the polarization $P(\{R\})$. 
In this situation the PES takes the form:
\begin{align}\label{eq:concl-1}
    E(\{R\}) = 
    E_{\textrm{DPLR}/\textrm{DP}}(\{R\}) -
    P(\{R\}) \cdot \mathcal E_\textrm{ext}(t)
\end{align}
%%           E[{R}] = E[{R}]_DPLR/DP - P[{R}] dot script[E(t)]  (x) 
%% We should use a different symbol for the energy (PES) and the  %electric field for which I use script E. Alternatively we may call the %PES Phi   
In Eq.~\eqref{eq:concl-1} the PES in absence of external field can be provided either by DPLR or by DP, depending on whether the size dependence of the electrostatic energy should be included explicitly or not. 
Eq.~\eqref{eq:concl-1} makes possible non-equilibrium MD simulations of an insulating system driven by an external electric field. The response of the system to the external field can also be studied with equilibrium MD using the Kubo formalism, as it was done in Ref.~\citen{zhang2020deep} to study the infrared absorption spectra of liquid water. 
In that work the standard DP model for the PES was used, which was adequate because at the infrared frequencies long-range electrostatic effects can be ignored as the dominant correlations are between neighboring molecular dipoles~\cite{chen2008role}. Long-range correlations among the molecular dipoles become important in the static ionic limit~\cite{ballenegger2004structure}.
In that limit, which is relevant to study the static dielectric constant of water, one should use the DPLR model for the PES in Eq.~\eqref{eq:concl-1}.              
 
Eq.~\eqref{eq:concl-1} can be further extended by including the quadratic response of the electrons via the susceptibility tensor $\chi$:
\begin{align}
    E(\{R\}) = 
    E_{\textrm{DPLR}/\textrm{DP}}(\{R\}) - 
    P(\{R\}) \cdot \mathcal E_\textrm{ext}(t) + 
    \frac12\mathcal E_\textrm{ext}(t) \cdot \chi(\{R\}) \cdot \mathcal E_\textrm{ext}(t)
\end{align}
%%   E[{R}] = E[{R}]_DPLR/DP - P[{R}] dot script[E] + 1/2 E(t) chi E(t) % (y)
This approach was used in the context of the Kubo formalism to compute the Raman spectra of water in Ref.~\citen{sommers2020raman}, a study that required the environmental dependence of the electric susceptibility.
The latter was modeled by a DNN extending the DW approach to the polarizability tensor. 
\recheck{The polarizability tensor is defined by the derivatives (i.e. the response) of the Wannier centroid positions with respect to an applied electric field in the limit of zero field. These derivatives are described by an extension of the DW model trained on DFT calculations in presence of a small, i.e. numerically infinitesimal, electric field, as discussed in Ref.~\citen{sommers2020raman}.}
The Raman scattering response is dominated by short range correlations between molecular polarizabilities, and the standard DP model was sufficient for the PES.

\section{Conclusions}

In this manuscript we examined a fundamental limitation of the local representation of the PES common to most ML models. These models are trained on quantum mechanical data, typically DFT data, on relatively small systems. They are then used in molecular simulations on larger systems that become accessible in view of the computational efficiency of the models. While successful in many cases, this procedure may lead to errors for physical properties affected by long-range electrostatic interactions. We have shown that these effects can be modeled accurately by the DPLR model, which extends the DP methodology using information on the centers of the electronic charge. The latter is encoded in the location of the Wannier centroids, which are averages of the Wannier centers uniquely asssociated to specific atoms. The environmental dependence of the Wannier centroids is described by DW, a DNN model, which we introduced in previous work to describe the dielectric polarization of insulators. Using the information from DW, DPLR augments the local representation of the PES of the standard DP model with the long-range electrostatic energy of ions and electrons, modeled by spherical Gaussian charge distributions centered at ions and Wannier centroids, respectively. The width of the Gaussian distributions is fixed by the value of the spread parameter $\beta$. Interestingly, the optimal $\beta$ that minimizes the generalization gap in the ML procedure, is close to the physical value expected from the spatial delocalization of the reference Wannier centroid distribution within DFT, as illustrated in our water and NaCl examples. The variation with size of the long-range electrostatic contribution, ignored in the standard DP model, is approximated accurately in the DPLR model, under the assumption that the electrostatic fields from which this contribution originates are a weak perturbation that can be treated within linear response theory. Because of that, it is possible to learn the environmental dependence of the Wannier centroids ignoring the size dependence of the electrostatic energy as done in the DW scheme.         
%\AZP{I think some comments on the width parameter $\beta$ and its selection should be added to the conclusions.}

DPLR uses two DNN models to represent the PES: the standard DP network for the short-range interactions and the DW network for computing long-range electrostatic interactions with the Ewald method.
Because of the complex architecture of the DPLR network, whereby force calculations require backward propagation within both DP and DW, and because of the need for Ewald calculations, DPLR simulations are more expensive than standard DP calculations.
Based on our experience, an increase of the computational burden of 
% at least one order of magnitude 
approximately a factor of 5 should be expected
when using DPLR in place of DP. The decision on which of the two models should be used in a given application should be guided by the physical properties of interest, whether long-range correlations originating by Coulomb forces are important or not. 

In developing the DPLR model we assumed that long-range electrostatic effects beyond the cutoff radius of the DP model originate a weak dependence on size that can be described within the linear response regime. This seems a reasonable assumption based on our experience and on the examples discussed in this manuscript. We cannot exclude, however, that effects of long-range electrostatics beyond the linear response regime may manifest in some circumstances. In these cases, the assumption made here that the environmental dependence of the Wannier centroids is not affected by size should be revisited, and one should envision a self-consistent condition for the centroids under the action of the long-range electrostatic field along lines similar to those developed in Ref.~\citen{gao2021self}. We expect that a better understanding of all these issues should emerge from widespread application of the DPLR model.

\recheck{
The major simplifying assumption of the current DPLR model is that the WCs are uniquely associated to specific atoms. Thus, they cannot split or recombine along molecular dynamics trajectories. This assumption limits the capability of the model to deal with chemical reactions: proton transfer reactions like those discussed, for example, in Ref.~\citen{zhang2020deep} and in Ref.~\citen{andrade2020free}, are allowed, but not, in general, electron transfer reactions, in which an electron is donated by one atomic entity to another. The centroid assumption shall be relaxed,
within the limits of adiabatic ground-state dynamics, in a future generalization of the DPLR model.  
}

\section*{Data availability}
\recheck{All data and codes needed to reproduce this work, including DP, DW, and DPLR models, are publicly available at \url{https://doi.org/10.5281/zenodo.6024644}.}

\section*{Acknowledgement}
This work was supported by the Computational Chemical Sciences Center ``Chemistry in Solution and at Interfaces" funded by the US Department of Energy under Award No. DE-SC0019394.  
The work of H.W.~was supported by the National Science Foundation of China under Grant No.11871110 and 12122103.

\appendix

\section{Modeling the short-range part}
\label{sec:app-srdp}

% The short-range part $E_\dpsr$ are the terms that decay rapidly with the distance between atoms, and is represented by the standard DP model, i.e.
In this appendix we briefly recall how the short-range DP model is constructed. The short-range part $E_\dpsr$ is given by a sum of atomic contributions  
\begin{align}\label{eq:dpsr}
    E_\dpsr = \sum_I E^\dpsr_I, \quad E^\dpsr_I = \fitting (\descrpt(\environ_I)),
\end{align}
where $\mathcal R_I$ is an environment matrix whose rows record the positions, relative to atom $I$, of the neighbors of that atom, within a specified cutoff radius $r_c$. We consider two sets of neighboring atoms $\mathcal N_I^a = \{R_J : \vert R_J - R_I\vert \leq r_c^a, J\neq I\}$ and  $\mathcal N_I = \{R_J : \vert R_J - R_I\vert \leq r_c, J\neq I\}$, where
$r_c^a$ and $r_c$ denote two cut-off radii, typically satisfying $r_c^a \leq r_c$. The number of neighbors of atom $I$ in the two sets are correspondingly denoted by $N^a(I)$ and $N(I)$. The two sets are used for constructing descriptors with angular and radial information.

The environment matrix $\tilde \environ_I^a$ is derived from the neighbor set $\mathcal N^a(I)$, and has dimension equal to $N^a(I)\times 4$, with each row being a four dimensional vector $s(R_{IJ}) \times (1, x_{IJ}/\vert R_{IJ}\vert, z_{IJ}/\vert R_{IJ}\vert, z_{IJ}/\vert R_{IJ}\vert )$. 
Here, the relative positions are defined as $R_{IJ} = R_I - R_J$, and their Cartesian components are denoted by $R_{IJ} = (z_{IJ}, y_{IJ}, z_{IJ})$.
$s(R_{IJ})$ is defined as $w(\vert R_{IJ}\vert) / \vert R_{IJ}\vert$, whith $w$ being a switch function that decays smoothly from 1 to 0 at the cut-off radius. 
% The environment matrix is the submatrix of $\tilde \environ_I$ taking all the rows and the last 3 columes.
The embedding matrix $\embedng_I^a$ has dimension equal to $N^a(I)\times M^a$, with each row mapping $s(R_{IJ})$ to  an $M^a$ dimensional vector via a feed forward DNN. A submatrix of $\embedng_I^a$ that includes its first $M_1$ columns ($M_1^a < M^a$) is denoted by $\embedng_I^{a,<}$.
The descriptor defined by $\descrpt^a = \frac1{(N^a(I))^2} (\embedng_I^{a,<})^T \tilde\environ_I^a (\tilde\environ_I^a)^T \embedng_I^a$ contains angular and radial information, and one proves that  it provides a symmetry preserving representation of the local environment of atom $I$.

The environment matrix $\tilde {\environ}_I$ includes only radial neighbor information. It is set up from the 
neighbor set $\mathcal N_I$ and has dimension equal to $N(I)\times 1$, with each row being $s(R_{IJ})$. 
The corresponding embedding matrix is denoted by $\embedng_I$ and has dimension equal to $N(I)\times M$. 
It should be noted that the embedding net used to map $s(R_{IJ})$ to $\embedng_I$ is usually different from the one used to generate $\embedng_I^a$. 
The descriptor with only radial information is defined as  $\descrpt^r = \frac{1}{N(I)} \sum_{j} (\embedng_I)_{jk}$, i.e., it is the average of the embedding matrix $\embedng_I$ with respect to its first index.
%RC: what does it mean summation of the first dimension of..???
The descriptor $\mathcal D^a$ contains radial and angular information on the environment, while $\mathcal D^r$ contains only radial information. 
Typically, combined angular and radial information is more important for close neighbors while radial information alone is sufficient to describe the more distant neighbors.
We thus define a hybrid descriptor as the concatenation of the radial and angular descriptors, i.e.~$\descrpt = (\descrpt^a, \descrpt^r)$.
Relative to the descriptor $\mathcal D = \mathcal D^a$, the hybrid descriptor $\descrpt = (\descrpt^a, \descrpt^r)$ has usually a stronger ability of generalization. 

% The descriptor is defined as the concatenation of the radial and angular descriptors, i.e.~$\descrpt = (\descrpt^a, \descrpt^r)$.
% The descriptor, denoted by $\descrpt$ in \eqref{eq:dpsr}, is defined by $\descrpt = (\embedng_I^<)^T \tilde\environ_I (\tilde\environ_I)^T \embedng_I$,
The hybrid descriptor is proved to be a symmetry preserving representation of the local environment of atom $I$.
$\fitting$ in \eqref{eq:dpsr} denotes the fitting network, which is implemented by a feed forward DNN with skip connections. 
The DNNs used in the standard DP construction are trained in an end-to-end way by stochastic gradient descent schemes like Adam~\cite{kingma2015}.

\end{document}